\documentclass[preprint2]{aastex6}

\usepackage{booktabs}
\usepackage{longtable}
\usepackage{xspace}
\usepackage[T1]{fontenc}
\usepackage{savesym} 
\savesymbol{tablenum} 
\usepackage{siunitx}
\restoresymbol{SIX}{tablenum} 

\usepackage{mathtools}

\newcommand{\RNum}[1]{\uppercase\expandafter{\romannumeral #1\relax}}

\newcommand{\psamp}{\ensuremath{\mathcal{P}}\xspace}
\newcommand{\ssamp}{\ensuremath{\mathcal{S}}\xspace}
\newcommand{\Kepler}{\textit{Kepler}\xspace} 
\newcommand{\isochrones}{\texttt{isochrones}\xspace}

\newcommand{\ntrial}{\ensuremath{n_\mathrm{trial}}\xspace}
\newcommand{\npl}{\ensuremath{n_\mathrm{pl}}\xspace}
\newcommand{\nnd}{\ensuremath{n_\mathrm{nd}}\xspace}
\newcommand{\ncell}{\ensuremath{n_\mathrm{cell}}\xspace}
\newcommand{\nstar}{\ensuremath{n_\star}\xspace}
\newcommand{\nssamp}{\ensuremath{n_{\ssamp}}\xspace}
\newcommand{\pval}{{\em p}-value\xspace}
\newcommand{\fcell}{\ensuremath{f_{\mathrm{cell}}}\xspace}
\newcommand{\nppass}{\ensuremath{n_{\mathrm{pl,pass}}}\xspace}
\newcommand{\nppassrun}{\ensuremath{n_{\mathrm{pl,pass,run}}}\xspace}
\newcommand{\fppassrun}{\ensuremath{f_{\mathrm{pl,pass,run}}}\xspace}
\newcommand{\nspass}{\ensuremath{n_{\star,\mathrm{pass}}}\xspace}
\newcommand{\nspassrun}{\ensuremath{n_{\star,\mathrm{pass,run}}}\xspace}
\newcommand{\fspassrun}{\ensuremath{f_{\star,\mathrm{pass,run}}}\xspace}
\newcommand{\dfdlogp}{\ensuremath{d f /d \log P}\xspace}
\newcommand{\smet}[1]{\ensuremath{M_{#1}\xspace}}
\newcommand{\ptr}{\ensuremath{p_{\mathrm{tr}}}}
\newcommand{\ptrr}[1]{\ensuremath{p_{\mathrm{tr},#1}}}
\newcommand{\pdet}{\ensuremath{p_{\mathrm{det}}}\xspace}
\newcommand{\pdett}[1]{\ensuremath{p_{\mathrm{det},#1}}\xspace}

\newcommand{\Mstar}{\ensuremath{M_{\star}}\xspace}
\newcommand{\Rstar}{\ensuremath{R_{\star}}\xspace} 
 
\newcommand{\fe}{\ensuremath{\mathrm{[Fe/H]}}\xspace}
\newcommand{\femn}{\ensuremath{\left<\mathrm{[Fe/H]}\right>}\xspace}
\newcommand{\teff}{\ensuremath{T_{\mathrm{eff}}}\xspace}  
\newcommand{\logg}{\ensuremath{\log g}\xspace} 
\newcommand{\vsini}{\ensuremath{v \sin i}\xspace} 
\newcommand{\kepmag}{\ensuremath{Kp}\xspace}

\newcommand{\Mp}{\ensuremath{M_{P}}\xspace} 
\newcommand{\Rp}{\ensuremath{R_P}\xspace}
\newcommand{\Rpp}[1]{\ensuremath{R_{P,#1}}\xspace}
\newcommand{\PP}[1]{\ensuremath{P_{#1}}\xspace}

\newcommand{\kms}{km s$^{-1}$\xspace}

\newcommand{\Me}{\ensuremath{M_{\oplus}}\xspace} 
\renewcommand{\Re}{\ensuremath{R_{\oplus}}\xspace}

\newcommand{\Rjup}{\ensuremath{R_{\mathrm{jup}}}\xspace}

\newcommand{\dex}{\ensuremath{\mathrm{dex}}\xspace}
\def\deg{\ensuremath{^{\circ}}}

\usepackage{xstring} 
\newcommand{\val}[1]{%
  \IfEqCase{#1}{%
{n-stars-cks}{1305}
{n-cand-cks}{2025}
{n-stars-lamost}{XX}
{n-stars-lamost-cut}{XX}
{n-cand-cks}{2025}
{f-pass-dilution}{0.922}%
{f-nopass-dilution}{7.8\%}%
{f-pass-spec}{0.969}%
{f-nopass-spec}{3.1\%}%
{b-cut}{0.9}%
  }[\PackageError{tree}{Undefined option to tree: #1}{}]%
}%

\newcommand{\fit}[1]{%
\IfEqCase{#1}{%
{per-all-se-logkp}{$-0.32^{+0.32}_{-0.25}$}
{per-all-se-beta}{$-0.3^{+0.2}_{-0.2}$}
{per-all-se-per0}{$6.5^{+1.6}_{-1.2}$}
{per-all-se-gamma}{$2.7^{+0.3}_{-0.2}$}
{per-all-se-gamma+beta}{$2.4^{+0.4}_{-0.3}$}
{per-all-sn-logkp}{$-0.28^{+0.18}_{-0.18}$}
{per-all-sn-beta}{$-0.1^{+0.1}_{-0.1}$}
{per-all-sn-per0}{$11.9^{+1.7}_{-1.5}$}
{per-all-sn-gamma}{$2.4^{+0.2}_{-0.2}$}
{per-all-sn-gamma+beta}{$2.3^{+0.2}_{-0.2}$}
{per-sub-se-logkp}{$-0.11^{+0.51}_{-0.39}$}
{per-sub-se-beta}{$-0.4^{+0.3}_{-0.4}$}
{per-sub-se-per0}{$8.8^{+2.3}_{-1.8}$}
{per-sub-se-gamma}{$3.0^{+0.4}_{-0.3}$}
{per-sub-se-gamma+beta}{$2.5^{+0.5}_{-0.4}$}
{per-sup-se-logkp}{$-0.35^{+0.32}_{-0.23}$}
{per-sup-se-beta}{$-0.4^{+0.2}_{-0.3}$}
{per-sup-se-per0}{$5.1^{+1.4}_{-0.8}$}
{per-sup-se-gamma}{$2.9^{+0.4}_{-0.3}$}
{per-sup-se-gamma+beta}{$2.5^{+0.5}_{-0.4}$}
{per-sub-sn-logkp}{$-0.75^{+0.27}_{-0.22}$}
{per-sub-sn-beta}{$+0.1^{+0.1}_{-0.2}$}
{per-sub-sn-per0}{$10.6^{+2.5}_{-1.8}$}
{per-sub-sn-gamma}{$2.7^{+0.4}_{-0.4}$}
{per-sub-sn-gamma+beta}{$2.8^{+0.5}_{-0.4}$}
{per-sup-sn-logkp}{$+0.09^{+0.29}_{-0.24}$}
{per-sup-sn-beta}{$-0.3^{+0.1}_{-0.2}$}
{per-sup-sn-per0}{$13.2^{+2.8}_{-2.1}$}
{per-sup-sn-gamma}{$2.5^{+0.2}_{-0.2}$}
{per-sup-sn-gamma+beta}{$2.1^{+0.2}_{-0.2}$}
{persmet-hot-se-logkp}{$-2.22^{+0.09}_{-0.09}$}
{persmet-hot-se-alpha}{$+1.7^{+0.1}_{-0.1}$}
{persmet-hot-se-beta}{$+0.6^{+0.2}_{-0.2}$}
{persmet-hot-sn-logkp}{$-2.95^{+0.15}_{-0.16}$}
{persmet-hot-sn-alpha}{$+2.2^{+0.2}_{-0.2}$}
{persmet-hot-sn-beta}{$+1.6^{+0.3}_{-0.3}$}
{persmet-hot-ss-logkp}{$-4.93^{+0.65}_{-0.76}$}
{persmet-hot-ss-alpha}{$+2.2^{+0.8}_{-0.7}$}
{persmet-hot-ss-beta}{$+5.5^{+1.6}_{-1.5}$}
{persmet-hot-jup-logkp}{$-3.32^{+0.29}_{-0.31}$}
{persmet-hot-jup-alpha}{$+0.9^{+0.4}_{-0.4}$}
{persmet-hot-jup-beta}{$+3.4^{+0.9}_{-0.8}$}
{persmet-warm-se-logkp}{$+0.05^{+0.22}_{-0.23}$}
{persmet-warm-se-alpha}{$-0.6^{+0.2}_{-0.2}$}
{persmet-warm-se-beta}{$-0.3^{+0.2}_{-0.2}$}
{persmet-warm-sn-logkp}{$-0.85^{+0.14}_{-0.13}$}
{persmet-warm-sn-alpha}{$+0.2^{+0.1}_{-0.1}$}
{persmet-warm-sn-beta}{$+0.5^{+0.2}_{-0.2}$}
{persmet-warm-ss-logkp}{$-2.59^{+0.53}_{-0.53}$}
{persmet-warm-ss-alpha}{$+0.5^{+0.3}_{-0.4}$}
{persmet-warm-ss-beta}{$+2.1^{+0.7}_{-0.7}$}
{persmet-warm-jup-logkp}{$-3.23^{+1.14}_{-1.23}$}
{persmet-warm-jup-alpha}{$+0.5^{+0.8}_{-0.8}$}
{persmet-warm-jup-beta}{$+0.8^{+1.4}_{-1.3}$}
}[\PackageError{tree}{Undefined option to tree: #1}{}]%
}%
\newcommand{\samp}[1]{%
\IfEqCase{#1}{%
{n-cand-cks}{2025}
{n-stars-cks}{1279}
{n-cand-cks-pass}{970}
{n-stars-cks-pass}{662}
{n-stars-cks-pass-smet<-0.4}{16}
{n-stars-field}{199991}
{n-stars-field-pass}{36959}
{n-stars-field-pass-eff}{33020}
{n-cks-lamo-tot}{476}
{n-cks-lamo-cal}{459}
{n-cks-lamo-out}{17}
{n-stars-lamo}{29997}
{n-stars-lamo-pass}{14382}
{lamo-smet-25}{-0.116}
{lamo-smet-50}{+0.020}
{lamo-smet-75}{+0.131}
{Hot Jupiters mean}{+0.19}
{Hot Jupiters pval}{\num{8e-04}}
{Hot Jupiters sem}{0.04}
{Hot Jupiters n}{14}
{Warm Jupiters mean}{+0.06}
{Warm Jupiters pval}{\num{6e-02}}
{Warm Jupiters sem}{0.02}
{Warm Jupiters n}{4}
{Cool Jupiters mean}{+0.06}
{Cool Jupiters pval}{\num{3e-01}}
{Cool Jupiters sem}{0.05}
{Cool Jupiters n}{13}
{Hot Sub-Saturns mean}{+0.26}
{Hot Sub-Saturns pval}{\num{7e-04}}
{Hot Sub-Saturns sem}{0.04}
{Hot Sub-Saturns n}{7}
{Warm Sub-Saturns mean}{+0.14}
{Warm Sub-Saturns pval}{\num{6e-05}}
{Warm Sub-Saturns sem}{0.02}
{Warm Sub-Saturns n}{19}
{Cool Sub-Saturns mean}{+0.11}
{Cool Sub-Saturns pval}{\num{1e-01}}
{Cool Sub-Saturns sem}{0.06}
{Cool Sub-Saturns n}{7}
{Hot Sub-Neptunes mean}{+0.11}
{Hot Sub-Neptunes pval}{\num{4e-09}}
{Hot Sub-Neptunes sem}{0.02}
{Hot Sub-Neptunes n}{108}
{Warm Sub-Neptunes mean}{+0.04}
{Warm Sub-Neptunes pval}{\num{2e-03}}
{Warm Sub-Neptunes sem}{0.01}
{Warm Sub-Neptunes n}{282}
{Cool Sub-Neptunes mean}{-0.05}
{Cool Sub-Neptunes pval}{\num{3e-01}}
{Cool Sub-Neptunes sem}{0.03}
{Cool Sub-Neptunes n}{29}
{Hot Super-Earths mean}{+0.05}
{Hot Super-Earths pval}{\num{2e-04}}
{Hot Super-Earths sem}{0.01}
{Hot Super-Earths n}{181}
{Warm Super-Earths mean}{-0.04}
{Warm Super-Earths pval}{\num{1e-01}}
{Warm Super-Earths sem}{0.02}
{Warm Super-Earths n}{132}
{Cool Super-Earths mean}{-0.36}
{Cool Super-Earths pval}{\num{2e-03}}
{Cool Super-Earths sem}{0.03}
{Cool Super-Earths n}{4}
{All Jupiters mean}{+0.12}
{All Jupiters pval}{\num{6e-04}}
{All Jupiters sem}{0.03}
{All Jupiters n}{31}
{All Sub-Saturns mean}{+0.16}
{All Sub-Saturns pval}{\num{1e-07}}
{All Sub-Saturns sem}{0.02}
{All Sub-Saturns n}{33}
{All Sub-Neptunes mean}{+0.05}
{All Sub-Neptunes pval}{\num{7e-07}}
{All Sub-Neptunes sem}{0.01}
{All Sub-Neptunes n}{419}
{All Super-Earths mean}{+0.01}
{All Super-Earths pval}{\num{8e-01}}
{All Super-Earths sem}{0.01}
{All Super-Earths n}{317}
{rate-hot-jup}{{0.57}^{+0.14}_{-0.12}}
{rate-hot-jup-simple}{{0.52}}
{L2-c0}{0.037}
{L2-c1}{-0.015}
{L2-c0-err}{0.002}
{L2-c1-err}{0.001}
{L2-val-m0.5}{+0.110 \pm 0.007}
{L2-val-p0.5}{-0.036 \pm 0.007}
{L1-c0}{0.033}
{L1-c1}{-0.015}
{L1-c0-err}{0.003}
{L1-c1-err}{0.001}
{pop-nstars}{100000}
{pop-nplanets}{110733}
{pop-intrate}{110.7}
}[\PackageError{tree}{Undefined option to tree: #1}{}]%
}%

\begin{document}

\title{The California-Kepler Survey. IV.\\ Metal-rich Stars Host a Greater Diversity of Planets}

\author{Erik A.\ Petigura\altaffilmark{1,7,8}}
\author{Geoffrey W.\ Marcy\altaffilmark{2}}
\author{Joshua N.\ Winn\altaffilmark{3}}
\author{Lauren M.\ Weiss\altaffilmark{4,9}}
\author{Benjamin J.\ Fulton\altaffilmark{1}}
\author{Andrew W.\ Howard\altaffilmark{1}}
\author{Evan Sinukoff\altaffilmark{5}}
\author{Howard Isaacson\altaffilmark{2}}
\author{Timothy D. Morton\altaffilmark{4}}
\author{John Asher Johnson\altaffilmark{6}}

\altaffiltext{1}{California Institute of Technology, Pasadena, CA 91125, USA}
\altaffiltext{2}{Department of Astronomy, University of California, Berkeley, CA 94720, USA}
\altaffiltext{3}{Princeton University, Princeton, NJ 08544, USA}
\altaffiltext{4}{University of Montreal, Montreal, QC, H3T 1J4, Canada}
\altaffiltext{5}{Institute for Astronomy, University of Hawai`i at M\={a}noa, Honolulu, HI 96822, USA}
\altaffiltext{6}{Harvard-Smithsonian Center for Astrophysics, Cambridge, MA 02138, USA}
\altaffiltext{7}{petigura@caltech.edu}
\altaffiltext{8}{Hubble Fellow}
\altaffiltext{9}{Trottier Fellow}

\begin{abstract}
Probing the connection between a star's metallicity and the presence and properties of any associated planets offers an observational link between conditions during the epoch of planet formation and mature planetary systems. We explore this connection by analyzing the metallicities of \Kepler target stars and the subset of stars found to host transiting planets. After correcting for survey incompleteness, we measure planet occurrence: the number of planets per 100 stars with a given metallicity $M$. Planet occurrence correlates with metallicity for some, but not all, planet sizes and orbital periods. For warm super-Earths having $P$ = 10--100~days and \Rp = 1.0--1.7~\Re, planet occurrence is nearly constant over metallicities spanning $-$0.4~dex to +0.4~dex. We find 20 warm super-Earths per 100 stars, regardless of metallicity. In contrast, the occurrence of warm sub-Neptunes (\Rp = 1.7--4.0~\Re) doubles over that same metallicity interval, from 20 to 40 planets per 100 stars. We model the distribution of planets as $d f \propto 10^{\beta M} d M$, where $\beta$ characterizes the strength of any metallicity correlation. This correlation steepens with decreasing orbital period and increasing planet size. For warm super-Earths $\beta$ = \fit{persmet-warm-se-beta}, while for hot Jupiters $\beta$ = \fit{persmet-hot-jup-beta}. High metallicities in protoplanetary disks may increase the mass of the largest rocky cores or the speed at which they are assembled, enhancing the production of planets larger than 1.7~\Re. The association between high metallicity and short-period planets may reflect disk density profiles that facilitate the inward migration of solids or higher rates of planet-planet scattering.
\end{abstract}


\section{Introduction}
\label{sec:intro}
Exploring the connection between planets and their host stars has been a long-standing focus of exoplanet astronomy. Host star metallicity is thought to reflect the metallicity of the protostellar nebula and the protoplanetary disk from which planets form. Metal-rich protoplanetary disks are thought to have enhanced surface densities of solids. Viewed in the context of core-accretion theory \citep{Lissauer95,Pollack96}, one might expect metal-rich disks to form terrestrial planets and the cores of gas giant planets with greater efficiency than metal-poor disks. If true, metal-rich stars should host greater numbers of gas giant and terrestrial planets. This prediction can be tested by studying the correlation (or lack thereof) between \fe and planet occurrence.

The extent to which stellar metallicity correlates with the presence or absence of planets has been the subject of many previous studies. \cite{Gonzalez97} observed that the first four extrasolar planets discovered orbited metal-rich stars and concluded that metal-rich stars have metal-rich protoplanetary disks which form planets more efficiently. As the sample of Doppler-detected planets grew into the hundreds, various studies noted a preference for Jovian-mass planets to orbit stars with super-solar metallicities (e.g. \citealt{Santos04,Fischer05}). However, as Doppler surveys pushed into lower regimes of planet mass, various authors noted that the correlation between planet occurrence and host star metallicity appeared to weaken (e.g. \citealt{Sousa08,Ghezzi10}). 

The prime \Kepler mission (2009--2013) revealed more than 4000 planet candidates with sizes as small as Mercury \citep{Barclay13}. In contrast to previous studies that explored the connection between host star metallicity and planet mass, \Kepler studies have focused mainly on the connection between metallicity and planet size. While a handful of \Kepler planets have well-measured masses through radial velocities (RVs; e.g., \citealt{Marcy14}) or transit-timing variations (TTVs; e.g., \citealt{Hadden17}), the vast majority of \Kepler planets have unknown masses. Current RV instruments require bright targets ($V < 13$~mag), and TTV measurements require tightly packed, multi-planet systems. Both TTV and RV mass measurements are only possible on a small fraction of the total \Kepler planet sample. 

While the \Kepler sample provided a large sample of planets for planet-metallicity studies, extensive follow-up spectroscopy was first required to measure host star metallicities. \cite{Buchhave12} measured the metallicities of 152 stars harboring 226 planets and observed that while planets larger than 4~\Re orbit metal-rich stars, smaller planets orbit stars with wide-ranging metallicities. Later, using an augmented sample of 405 stars hosting 600 planets, \cite{Buchhave14} argued for three distinct stellar metallicity distributions, with breakpoints at 1.7~\Re and 3.9~\Re. In contrast, \cite{Schlaufman15} found no evidence for different metallicity distributions above and below 1.7~\Re and raised several concerns regarding the statistical validity of the \cite{Buchhave14} analysis.

RV mass measurements of transiting planets have revealed that planets smaller than 1.7~\Re have bulk densities consistent with rocky compositions (e.g. \citealt{Weiss14,Rogers15}), at least for the short-period planets ($P \lesssim 20$~days) that are amenable to these RV mass measurements. Consequently, the degree to which metallicity correlates with the occurrence of planets smaller than 1.7~\Re is of particular interest. On this point, there is disagreement in the literature. \cite{Wang15c}, using low precision metallicities from the Kepler Input Catalog (KIC; \citealt{Brown11}), reported that the rate of planets smaller than 1.7~\Re is $1.72^{+0.18}_{-0.17}$ times higher for stars with \fe > 0~dex compared to stars \fe < 0~dex. In contrast, \cite{Buchhave15} found no evidence of a metallicity enhancement among stars hosting planets smaller than 1.7~\Re.

A long-standing limitation of \Kepler metallicity studies was that reliable metallicities did not exist for a representative number of \Kepler field stars. In constructing the KIC, \cite{Brown11} placed a prior on the \Kepler field metallicities based on the metallicities of nearby stars, as measured by \cite{Nordstrom04}, which have a mean of $-0.14$~dex and dispersion of 0.19~dex. However, it was not clear whether \Kepler field stars, which are typically $\sim$1~kpc from Earth, would follow the metallicity distribution  of nearby stars.

Such metallicity offsets have been invoked to explain differences in planet occurrence rates between the \Kepler field and the solar neighborhood. \cite{Howard12} measured a hot Jupiter occurrence rate of $0.4\pm0.1\%$, roughly 40\% that in the solar neighborhood ($1.2\pm0.4\%$, \citealt{Wright12}), and speculated that ``a paucity of metal-rich stars in the \Kepler sample is one possible explanation.''

Today we know that \Kepler field stars have a {\em higher} mean metallicity than solar neighborhood stars. The Large Sky Area Multi-object Fibre Spectroscopic Telescope (LAMOST; \citealt{Zhao12,Cui12,Luo12}) is instrumented with a multiplexed, low-resolution spectrometer (4000 fibers, $R$ = 1800) and can therefore efficiently gather spectra for large samples of \Kepler target stars with and without transiting planets.  \cite{Dong14}, through analyzing the LAMOST Data Releases 1 and 2, found that the mean metallicity of 12,000 \Kepler field stars was $-0.04$~dex, much closer to solar than to the value of $-0.14$~dex assumed in the construction of the KIC. \cite{Guo17} also found a near-solar mean metallicity of $-0.04$~dex in an analysis of 610 \Kepler field stars observed by the Hectochelle $R = 34,000$ spectrometer at the MMT. Thus, a metallicity offset cannot explain differences in the hot Jupiter rates.

The LAMOST datasets permitted several breakthroughs in \Kepler planet-metallicity studies, by measuring the true metallicity distribution of bright (\kepmag < 14) \Kepler field stars. \cite{Mulders16} analyzed the LAMOST metallicities and a sample of 665 planet candidates, and found that occurrence rate of hot small planets ($P$ < 10~days, \Rp < 4~\Re) is three times higher among super-solar metallicity hosts compared to sub-solar hosts. \cite{Dong17} also analyzed LAMOST metallicities and a sample of 295 planets and reported a similar trend, also noting that hot Neptune-size planets are typically single.

Here, we examine the connection between planets and stellar metallicity using spectroscopy from the California-Kepler Survey (CKS). Given that the CKS produced a homogeneous set of highly precise metallicities of \val{n-stars-cks} stars, we can explore this connection in unprecedented detail. A key advantage of the CKS sample is the highly-precise planet radii (10\% precision) and stellar metallicities (0.04~dex precision) compared to previous studies. The CKS sample has high purity (i.e. low false positive rate) due to extensive vetting of false positives. This dataset has already revealed new features in the planet radius distribution, most notably a gap in the size distribution of small (\Rp~=~1--4~\Re planets) reported by \cite{Fulton17}.

We find that planets larger than Neptune are preferentially found around metal-rich stars, while planets smaller than Neptune are found around stars of wide-ranging metallicity. For super-Earth-size planets (\Rp = 1.0--1.7~\Re), we observe a positive metallicity correlation for $P$~=~1--10~days, but no correlation for $P$ = 10--100~days. In contrast, rates of sub-Neptune-size planets (\Rp = 1.7--4.0~\Re) correlate with metallicity over $P$ = 1--100~days. Planets larger than Neptune are an order of magnitude less common than planets smaller than Neptune, and probing the possible metallicity correlations is more challenging. However, we observe strong metallicity correlations for both Jovian-size (\Rp = 8.0--24.0~\Re) and sub-Saturn-size (\Rp = 4.0--8.0~\Re) planets.

We describe our planet and stellar samples in Section~\ref{sec:sample}. Section~\ref{sec:smet} explores the distribution of planets in the $P$--\Rp plane as a function of metallicity, and highlights areas where metallicity plays a strong effect. In Sections~\ref{sec:occur-methodology}--\ref{sec:occur-per-radius-smet}, we compare host star metallicities to the \Kepler field star distribution and compute planet occurrence as a function of metallicity. We summarize our findings in Section~\ref{sec:discussion}, and offer some interpretations of the observed trends.

\section{Sample}
\label{sec:sample}
Studies of  planet occurrence require both a sample of planets \psamp and a parent stellar sample \ssamp from which the planets are drawn. We construct \psamp from the CKS sample (Section~\ref{sec:sample-planet-initial}) and apply a series of filters aimed at creating a well-defined sample of high purity with well-measured radii (Section~\ref{sec:sample-planet-filter}). We then construct \ssamp from the \Kepler field stars after applying the same set of filters used to construct \psamp (Section~\ref{sec:sample-field}). Because the metallicities are not known for every star in \ssamp, we employ the LAMOST parameters as a proxy (Section~\ref{sec:sample-lamost}).

\subsection{Initial Planet Sample}
\label{sec:sample-planet-initial}
CKS is a large-scale spectroscopic survey of $\val{n-stars-cks}$ \Kepler Objects of Interest (KOIs). The sample selection, spectroscopic observations, and spectroscopic analysis are described in detail in \citet[hereafter Paper~\RNum{1}]{Petigura17b}. In brief, the sample was initially constructed by selecting all \Kepler Objects of Interest (KOIs) brighter than $Kp$ = 14.2~mag.%
\footnote{\Kepler magnitude; $\kepmag \approx V - 0.4$~mag for G2 stars.}
A KOI is a \Kepler target star which showed periodic photometric dimmings indicative of planet transits. However, not all KOIs have received the necessary follow-up attention needed to confirm the planets.

Over the course of the project, we included additional targets to cover different planet populations, including multi-candidate hosts, Ultra Short Period candidates, and Habitable Zone candidates. We obtained spectra at the 10~m Keck Telescope using the High Resolution Echelle Spectrometer at a resolution of $R = 60,000$ \citep{Vogt94}. A key feature of the CKS dataset is that stars were observed to a consistent signal-to-noise ratio (S/N) of 45/pixel$^{-1}$ at the peak of the blaze function near 5500~\AA.

In Paper~\RNum{1}, we extracted the following properties for each star: effective temperature \teff, surface gravity \logg, metallicity \fe, and projected rotational velocity \vsini using the SpecMatch \citep{Petigura15thesis} and SME@XSEDE codes (Paper~\RNum{1}). 

In \citet[hereafter Paper~\RNum{2}]{Johnson17}, we converted the spectroscopic properties of \teff, \logg, and \fe into stellar mass \Mstar, radius \Rstar, and age. This conversion is facilitated by the publicly available  \isochrones Python package \citep{Morton15}. Stellar mass, radius, and age are measured to 4\%, 11\%, and 30\%, respectively. We then recompute planetary radii and equilibrium temperatures using our updated CKS parameters and the results from transit fitting performed by \cite{Mullally15}. Paper~\RNum{2} provides updated stellar and planetary properties for \val{n-stars-cks} KOIs and \val{n-cand-cks} planet candidates, which are the starting point for our planet sample \psamp.
 	
\subsection{Filtered Planet Sample}
\label{sec:sample-planet-filter}
We applied a series of filters, described below in bullets, to the CKS sample to arrive at our final planet sample \psamp. The filters restrict the range of stellar properties included in our analysis. Since we aim to explore the connection between metallicity and planet size, we also require the planets have well-determined radii.

Where possible, we filter based on the DR25 stellar properties table of \cite{Mathur17}. These cuts may be applied homogeneously to the field star sample \ssamp. Some filters involve follow-up observations not available for every star in \ssamp. In Section~\ref{sec:sample-field}, we quantify the number of field stars that would have been excluded had comparable follow-up been performed. The applied filters are as follows:

\begin{enumerate}
\item {\em Stellar brightness.} We restrict our sample to the magnitude-limited sub-sample of the CKS sample (i.e. $Kp < 14.2$~mag). 

\item {\em Stellar effective temperature}. The spectroscopic tools used in Paper~\RNum{1} produce reliable results for \teff = 4700--6500~K. We restrict our analysis to stars having photometric \teff = 4700--6500~K (DR25 stellar properties table).

\item {\em Stellar surface gravity}. We restrict our analysis to stars having photometric \logg = 3.9--5.0~dex (DR25 stellar properties table).

\item {\em Stellar dilution}. Dilution from nearby stars can also alter the apparent planetary radii. \cite{Furlan17} compiled high resolution imaging observations performed by several groups.%
\footnote{Alphabetical by author: \cite{Adams12,Adams13,Baranec16,Cartier15,Dressing14,Everett15,Gilliland15,Horch12,Horch14,Howell11,Law14,Lillo-Box12,Lillo-Box14,Wang15a,Wang15b,Ziegler17}.}
When a nearby star is detected, \cite{Furlan17} computed a radius correction factor (RCF), which accounts for dilution assuming the planet transits the brightest star. We elect to not apply this correction factor, but conservatively exclude KOIs where the RCF is larger than 5\%. 

\item {\em Planet orbital period}. We remove planet candidates with orbital periods longer than 350~days. This excludes planets that only transit once or twice during \Kepler observations, which have a higher false positive rate \citep{Mullally15}. 

\item {\em Planet false positive designation}. We exclude candidates that are identified as false positives according to Paper~\RNum{1}.

\item {\em Planets with grazing  transits}. Finally, we exclude stars having grazing transits ($b$ > \val{b-cut}), which have suspect radii due to covariances with the planet size and stellar limb-darkening during the light curve fitting.
\end{enumerate}

Our successive filters, along with a running tally of the number planets which pass them, are listed in Table~\ref{tab:cks-cuts}. The filters on dilution and grazing transits each eliminate a small percentage (8\% and 5\%) of planets with imprecise radii, even after CKS spectroscopy. Similar filters clarified the bimodal planet radius distribution presented by \cite{Fulton17}, Paper~\RNum{3} in the CKS series.

\begin{figure*}
\centering
\includegraphics[width=\textwidth]{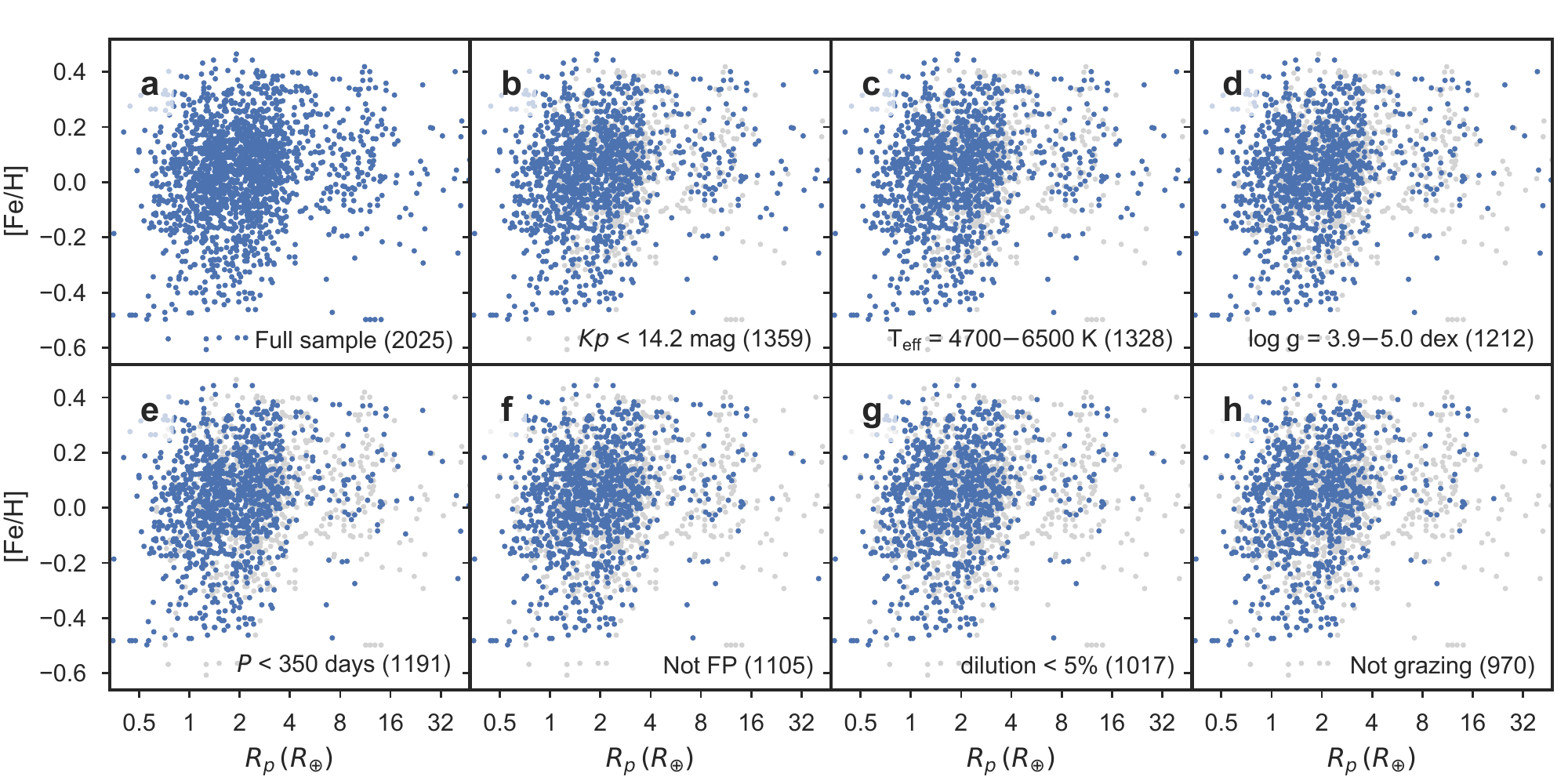}
\caption{Host star metallicity vs. planet size after the application of successive filters. Panel (a): all KOIs in the CKS sample. Panels (b)--(j): blue points show the KOIs that pass successive filters in stellar brightness, effective temperature, surface gravity, planetary orbital period, false positive disposition, radius precision, dilution due to nearby stars, and impact parameter. The gray points show KOIs that do not pass one or more of the filters.}
\label{fig:prad-smet-cuts}
\end{figure*}

In total, \psamp contains \samp{n-cand-cks-pass} planets orbiting \samp{n-stars-cks-pass} stars that pass all filters. Figure~\ref{fig:prad-smet-cuts} shows planets on the \fe--\Rp plane as successive cuts are applied. Even though planets across this plane are filtered out, planets larger than Neptune have a higher rate of removal. Properties of the stars hosting these planets are shown in Figure~\ref{fig:cks-lamo-cuts}.

\begin{deluxetable*}{lrrrr}
\tablecaption{Filters Applied to Planet Sample\label{tab:cks-cuts}}
\tablecolumns{5}
\tablehead{
\colhead{Filter} & 
\colhead{\nppass} & 
\colhead{\nppassrun} & 
\colhead{\fppassrun} &
\colhead{\nspassrun} 
}
\tablewidth{0pt}
\startdata
Full sample                         & 2025  & 2025  & 1.000 & 1279  \\
$Kp$ < 14.2 mag                     & 1359  & 1359  & 0.671 & 954   \\
$\teff$ = $4700-6500$ K             & 1977  & 1328  & 0.977 & 929   \\
$\logg$ = $3.9-5.0$ dex             & 1899  & 1212  & 0.913 & 829   \\
$P$ < 350 days                      & 1965  & 1191  & 0.983 & 817   \\
Not a false positive                & 1861  & 1105  & 0.928 & 758   \\
Radius correction factor < 5\%      & 1891  & 1017  & 0.920 & 695   \\
Not a grazing transit ($b$ < 0.9)   & 1870  & 970   & 0.954 & 662   \\
\enddata
\tablecomments{Summary of the filters applied to the CKS catalog to create planet sample \psamp. The column labeled \nppass is the total number of KOIs that pass a specific filter and \nppassrun is a running tally of KOIs that pass all filters. For filter $i$, $\fppassrun(i) = \nppassrun(i)/\nppassrun(i-1)$. For example, $\fppassrun(3)$ = $\nppassrun(3)/\nppassrun(2) = 1328/1359 = 0.977$. The column labeled \nspassrun gives the numbers of unique stars hosting the \nppassrun KOIs.}
\end{deluxetable*}

\subsection{Field Star Sample}
\label{sec:sample-field}
In order to compare the properties of stars with and without transiting planets, we need to consider the parent population of \Kepler field stars \ssamp, from which \psamp is drawn. We begin with the \cite{Mathur17} catalog that lists all \samp{n-stars-field} stars observed at some point during the \Kepler mission. Where possible, we apply the same set of filters to construct \ssamp that were used to construct \psamp. Table~\ref{tab:field-cuts} summarizes the number of stars that pass cuts on \kepmag, photometric \teff, and photometric \logg. A total of \samp{n-stars-field-pass} stars pass all cuts. The distribution of field star properties is shown in Figure~\ref{fig:cks-lamo-cuts}.

The filters used to construct \psamp were not simply cuts on the stellar properties, but also on quality of the CKS stellar radii and the properties of the planet candidates. We cannot apply these same filters to the field stars on a star-by-star basis because these stars do not have CKS stellar radii or detected planets. However, we must assess whether the remaining filters from Section~\ref{sec:sample-planet-filter} would have excluded a significant fraction of field stars, assuming all stars received similar follow-up attention. Here, we quantify the number of field stars that would have been excluded from \ssamp had comparable follow-up been performed.

\begin{enumerate}
\item {\em Reliable spectroscopic parameters.} Our initial sample of planets orbit stars for which the spectroscopic analysis described in Section~\ref{sec:sample} produced reliable results. For a star to be included, \vsini must be less than 20~\kms. This selection effect excludes some stars that are typically near 6500~K, where \vsini values begin to exceed 20~\kms. After filtering on \kepmag, photometric \teff, and photometric \logg, 3.1\% of stars are excluded because they do not have reliable spectroscopic parameters.

\item {\em Stellar dilution}. 7.8\% of planet candidates were excluded because the \Kepler apertures contained enough flux from neighboring stars, such that the planet radii required a correction factor larger than 5\%. The fraction of stars excluded depends on the crowding of \Kepler field stars. We make the assumption that this distribution is independent of KOI status. Had all field stars received a similar level of high-contrast imaging follow-up, 7.8\% would have companions sufficiently bright and close to warrant a RCF of $>5\%$.

\item {\em Planet orbital period}. For a KOI to be included in our sample, we required that the orbital period be less than 350~days. Such a cut is strictly a cut on the planet properties and does not preclude any stars from being included in \ssamp.

\item {\em Planet false positive disposition}. A star must first be identified as a KOI in order to then be designated as a false positive. Therefore, the false positive filter does not exclude significant numbers of field stars from \ssamp.

\item {\em Planets with grazing transits}. We require that the planets have non-grazing  impact parameters. As with the radius filter, it does not affect inclusion in \ssamp. However, this cut on impact parameter must be factored into the geometric transit probability, which affects the occurrence calculation described in Section~\ref{sec:occur-methodology}.
\end{enumerate}

When we account for stars that would have been excluded from our sample due to unreliable spectroscopic parameters or the presence of nearby stars, we find that the planet sample \psamp was drawn from a parent stellar sample \ssamp containing $\samp{n-stars-field-pass} \times \val{f-pass-dilution}   \times \val{f-pass-spec} = \samp{n-stars-field-pass-eff}$ stars.

\begin{deluxetable*}{lrrrh}
\tablecaption{Filters Applied to Stellar Sample\label{tab:field-cuts}}
\tablecolumns{4}
\tablehead{
\colhead{Cut} & 
\colhead{\nspass} & 
\colhead{\nspassrun} & 
\colhead{\fspassrun} &
\nocolhead{\nspassrun}
}
\tablewidth{0pt}
\startdata
Full sample                         & 199991 & 199991 & 1.000 & 199991 \\
$Kp$ < 14.2 mag                     & 81758 & 81758 & 0.409 & 81758 \\
$\teff$ = $4700-6500$ K             & 168885 & 62751 & 0.768 & 62751 \\
$\logg$ = $3.9-5.0$ dex             & 162854 & 36959 & 0.589 & 36959 \\
\enddata
\tablecomments{Summary of the filters applied to \ssamp. See Table~\ref{tab:cks-cuts} for column descriptions.}
\end{deluxetable*}

\subsection{Metallicity Distribution of Kepler Field Stars}
\label{sec:sample-lamost}
Here, we characterize the metallicity distribution of the parent sample \ssamp. While the KIC \citep{Brown11} and its updates (e.g. \citealt{Mathur17}) tabulate metallicities for every \Kepler target star, they are inadequate for characterizing the metallicity distribution of the \Kepler field given their low precision. Instead, we use the LAMOST/LASP stellar parameters \citep{Luo15} from Data Release 3 (DR3).%
\footnote{\url{http://dr3.lamost.org}}
Following \cite{De-Cat15}, we crossmatch the LAMOST-DR3 with the KIC by identifying LAMOST spectra taken within 1.2~arcsec of a KIC target star. This crossmatching results in \samp{n-stars-lamo} stars in common having \kepmag < 14.2~mag. We note a systematic offset between the LAMOST and CKS metallicity scales of $\approx$0.04~dex when comparing \samp{n-cks-lamo-tot} stars in common. We apply a correction, described in Appendix~\ref{sec:calibrate-lamost}, but estimate that residual systematic offsets of $\approx$0.01~dex remain.

\begin{figure*}
\centering
\includegraphics[width=\textwidth]{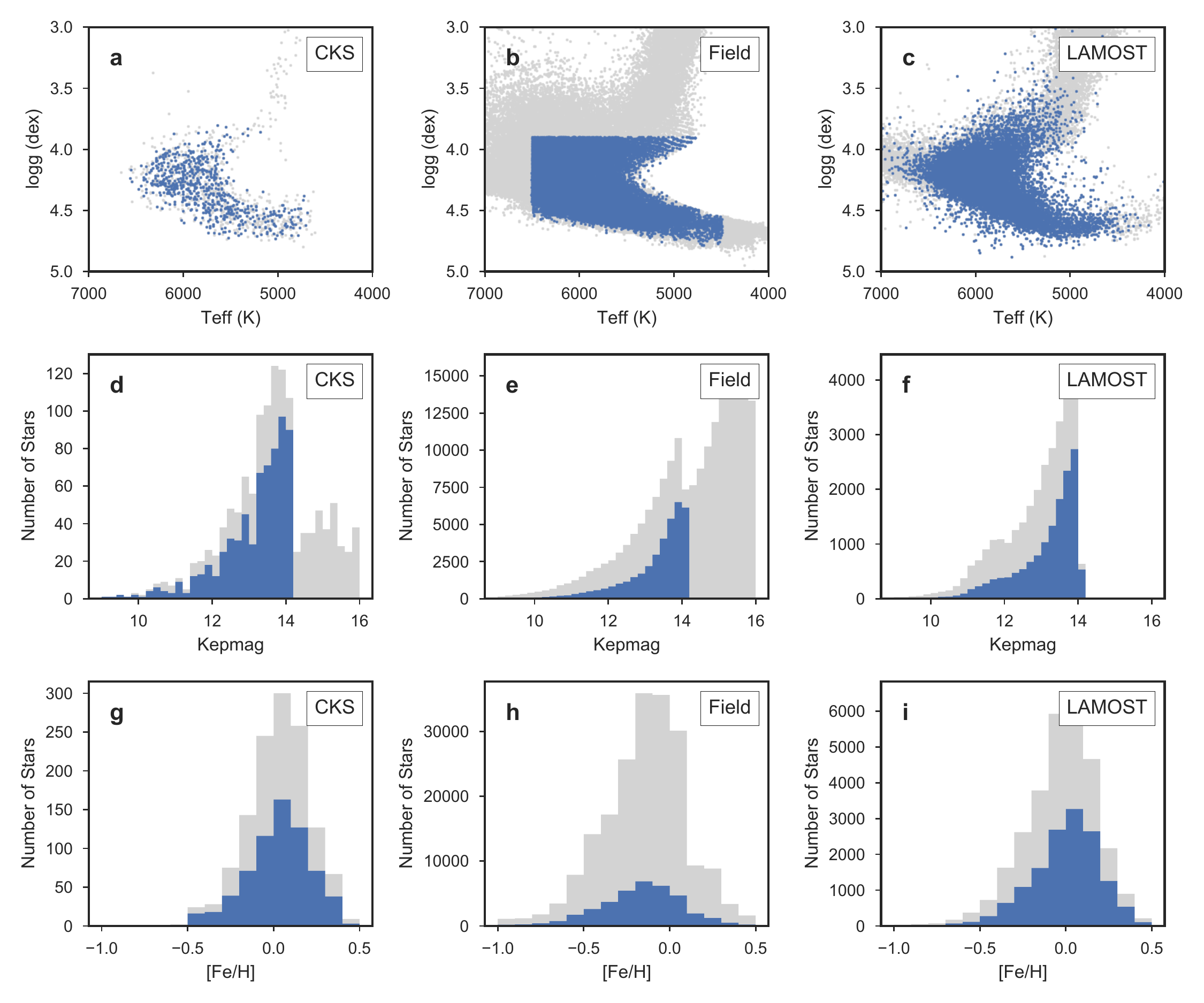}
\caption{The distribution of stellar properties for the three samples of stars considered in this work. Panel (a): blue points show \teff and \logg of CKS planet hosts that passed all filters \psamp; gray points represent all CKS planet hosts. Panels (b) and (c) show the same quantities as panel (a), but for the \Kepler target stars and the LAMOST sample, respectively.  Panels (d)--(f): Distributions of  host star \kepmag. Panels (g)--(i): distributions of host star metallicity from different catalogs. The sub-solar mean metallicity of the \Kepler target stars (h) is a reflection of the low metallicity prior used in the photometric modeling.\label{fig:cks-lamo-cuts}}
\end{figure*}

We apply the same set of filters to the LAMOST stars that we applied to \psamp. These include cuts in \kepmag, \teff, and \logg. After applying these cuts, we are left with \samp{n-stars-lamo-pass} stars with LAMOST parameters. The filters are summarized in Table~\ref{tab:lamo-cuts} and the distribution of LAMOST stellar properties is shown in Figure~\ref{fig:cks-lamo-cuts}.

Table~\ref{tab:smet-statistics} summarizes the metallicity distribution of \psamp and \ssamp measured through different methods. The mean metallicity of \ssamp, as measured by LAMOST, is $-$0.01~dex, similar to the mean metallicity of the filtered planet sample \psamp, $+$0.03~dex, as measured by CKS.  We note that the mean metallicity of \ssamp, according to the DR25 stellar properties table \citep{Mathur17}, is $-0.19$~dex. This low value is due to the low metallicity prior used in the photometric modeling.

\begin{deluxetable*}{lrrrh}
\tablecaption{Summary of Cuts to LAMOST Sample\label{tab:lamo-cuts}}
\tablecolumns{4}
\tablehead{
\colhead{Cut} & 
\colhead{\nspass} & 
\colhead{\nspassrun} & 
\colhead{\fspassrun} &
\nocolhead{\nspassrun}
}
\tablewidth{0pt}
\startdata
Full sample                         & 29997 & 29997 & 1.000 & 29997 \\
$Kp$ < 14.2 mag                     & 29997 & 29997 & 1.000 & 29997 \\
$\teff$ = $4700-6500$ K             & 23551 & 23551 & 0.785 & 23551 \\
$\logg$ = $3.9-5.0$ dex             & 18625 & 14382 & 0.611 & 14382 \\
\enddata
\tablecomments{Summary of the cuts applied to the LAMOST DR-2 sample.  See Table~\ref{tab:cks-cuts} for column descriptions.}
\end{deluxetable*}

\begin{deluxetable*}{lRRR}
\tablecaption{Comparison of Planet Host and Field Star  Metallicities\label{tab:smet-statistics}}
\tablecolumns{4}
\tablehead{
\colhead{} & 
\colhead{\psamp (spec)} &
\colhead{\ssamp (phot)} &
\colhead{\ssamp (spec)} \\
\colhead{} & 
\colhead{(dex)} & 
\colhead{(dex)} & 
\colhead{(dex)} 
}
\tablewidth{0pt}
\startdata
Mean & 0.031 & -0.187 & -0.005 \\
RMS & 0.185 & 0.259  & 0.207 \\
SEM & 0.006 & 0.001 & 0.002 \\
25\% & -0.069 & -0.320 & -0.116 \\
50\% & 0.052 & -0.160 & 0.020 \\
75\% & 0.148 & -0.020 & 0.131 \\
\enddata
\tablecomments{Summary of metallicity distributions for different stellar samples. \psamp (spec) refers to the CKS metallicities of our filtered list of planet hosts (Section~\ref{sec:sample-planet-filter}). \ssamp (phot) refers to the photometric metallicities of our parent stellar sample (Section~\ref{sec:sample-field}). \ssamp (spec) refers to the metallicities of \ssamp measured using LAMOST data (Section~\ref{sec:sample-lamost}). \ssamp (phot) and \ssamp (spec) have significantly different mean metallicities, due the metallicity prior imposed by \cite{Mathur17}. }
\end{deluxetable*}

\section{Metallicities of Planet Hosts}
\label{sec:smet}
In this section, we examine the metallicities of planet host stars with respect to planet size and orbital period. We first briefly note the trends seen within the planet sample \psamp, without reference to field star metallicities (Section~\ref{sec:smet-planets}). However, features in $P$--\Rp--\fe distribution are much more apparent when compared to the metallicities of the parent stellar sample \ssamp. We perform this analysis in Section~\ref{sec:smet-planets-field}.  By searching for evidence of elevated metallicity in the host stars of certain types of planets, we can identify planet classes that show some positive correlation with metallicity. The advantage of this approach is that it does not require any modeling of the \Kepler survey completeness. A more difficult but ultimately more useful approach is to compute planet occurrence as a function of metallicity, which requires modeling survey completeness. This approach is taken in Section~\ref{sec:occur-per-radius-smet}.

\subsection{Properties of Planet Hosts}
\label{sec:smet-planets}
\begin{figure*}
\includegraphics[width=\textwidth]{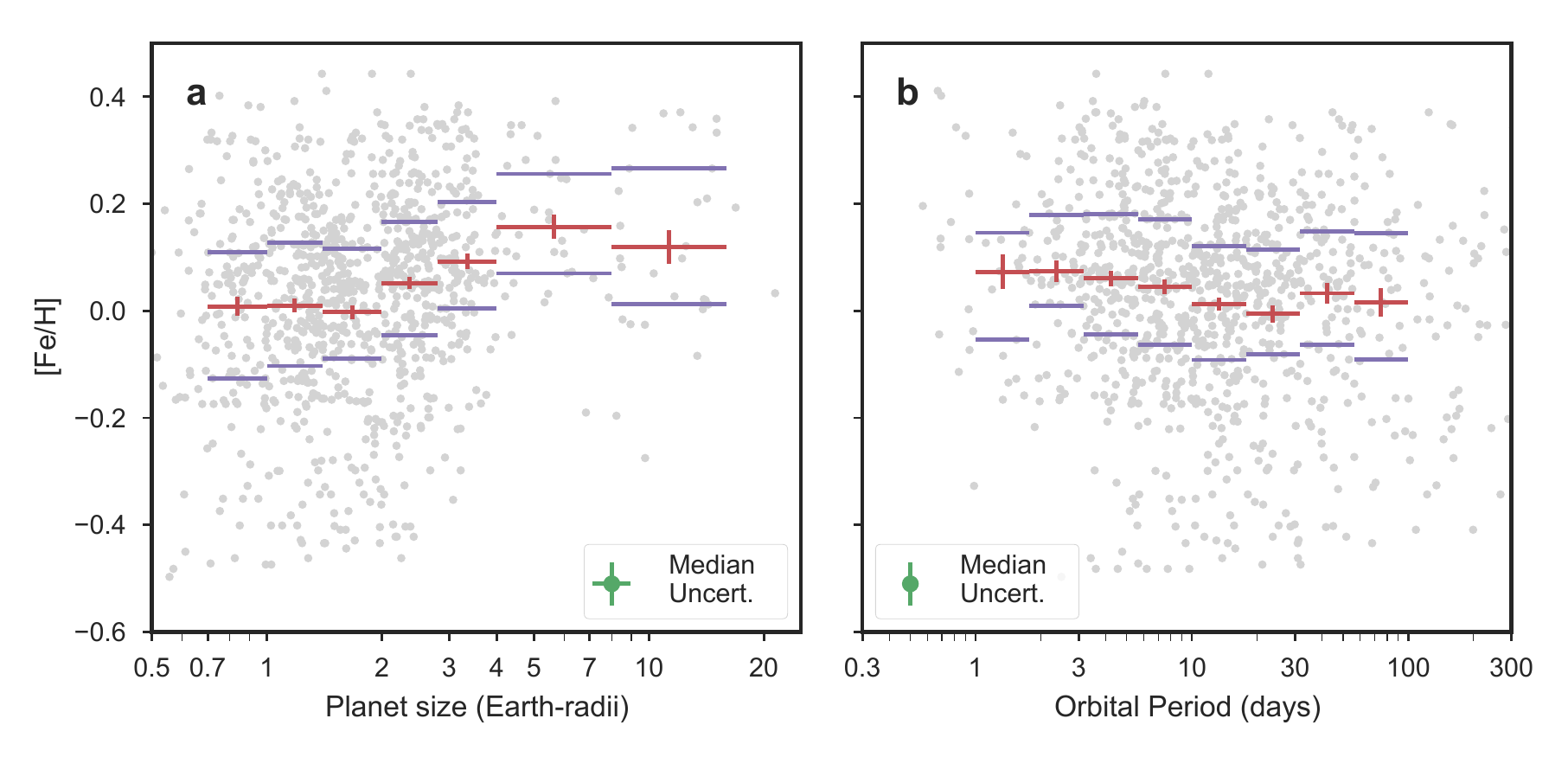}
\caption{Panel (a) shows the sizes and host star metallicities for the $\samp{n-cand-cks-pass}$ planets in the filtered CKS planet catalog \psamp. We observe a clear deficit of planets larger than Neptune with sub-Solar metallicities. The \cite{Fulton17} radius gap may be seen from \Rp = 1.5--2.0~\Re. We show the mean host star metallicity for various size ranges of planets with the red lines. The vertical bars show the standard error of the mean. The purple lines show the 25\% and 75\% quantiles. Mean metallicity is roughly constant from 0.7~\Re to 2.0~\Re, rises from 2.0~\Re to 4.0~\Re, and is roughly constant from 4~\Re to 16~\Re. Panel (b): same as (a) except showing orbital period on the x-axis. We observe a small 0.05~dex increase in mean metallicity with decreasing orbital period over $P$ = 1--10~days.\label{fig:prad-fe}}
\end{figure*}

In Figure~\ref{fig:prad-fe}, we show the planet sizes and host star metallicities for the $\val{n-cand-cks}$ planet candidates in the CKS sample. Planets smaller than Neptune are found around stars of wide-ranging metallicities, while there is a  deficit of planets larger than Neptune around stars with sub-solar metallicity. 

To investigate variation in average host star metallicities with planet sizes, we divided the metallicity measurements according to planet size. Bins span a factor of $\sqrt{2}$ in \Rp for planets smaller than 4~\Re. Larger planets are placed into bins spanning a factor of 2, owing to lower numbers. Figure~\ref{fig:prad-fe} shows the mean metallicities for these planet radius bins along with the 25\% and 75\% quantiles. We observe a gradual upward trend in mean host star metallicity from smaller to larger planets. \cite{Buchhave12,Buchhave14} observed a similar trend in smaller samples of planet hosts.

Figure~\ref{fig:prad-fe} also shows host star metallicity as a function of orbital period. Unlike the \Rp--\fe, there are no large regions of the $P$--\fe plane clearly devoid of planets. After computing the mean metallicity in bins of $P$ spanning 0.25~dex, we observe a slight increase in mean metallicity of about 0.05~dex with decreasing orbital period over $P$ = 1--10~days. \cite{Mulders16} observed a similar trend in a smaller sample of planet hosts.

\subsection{Comparison to Field Stars}
\label{sec:smet-planets-field}
\begin{figure*}
\centering
\includegraphics{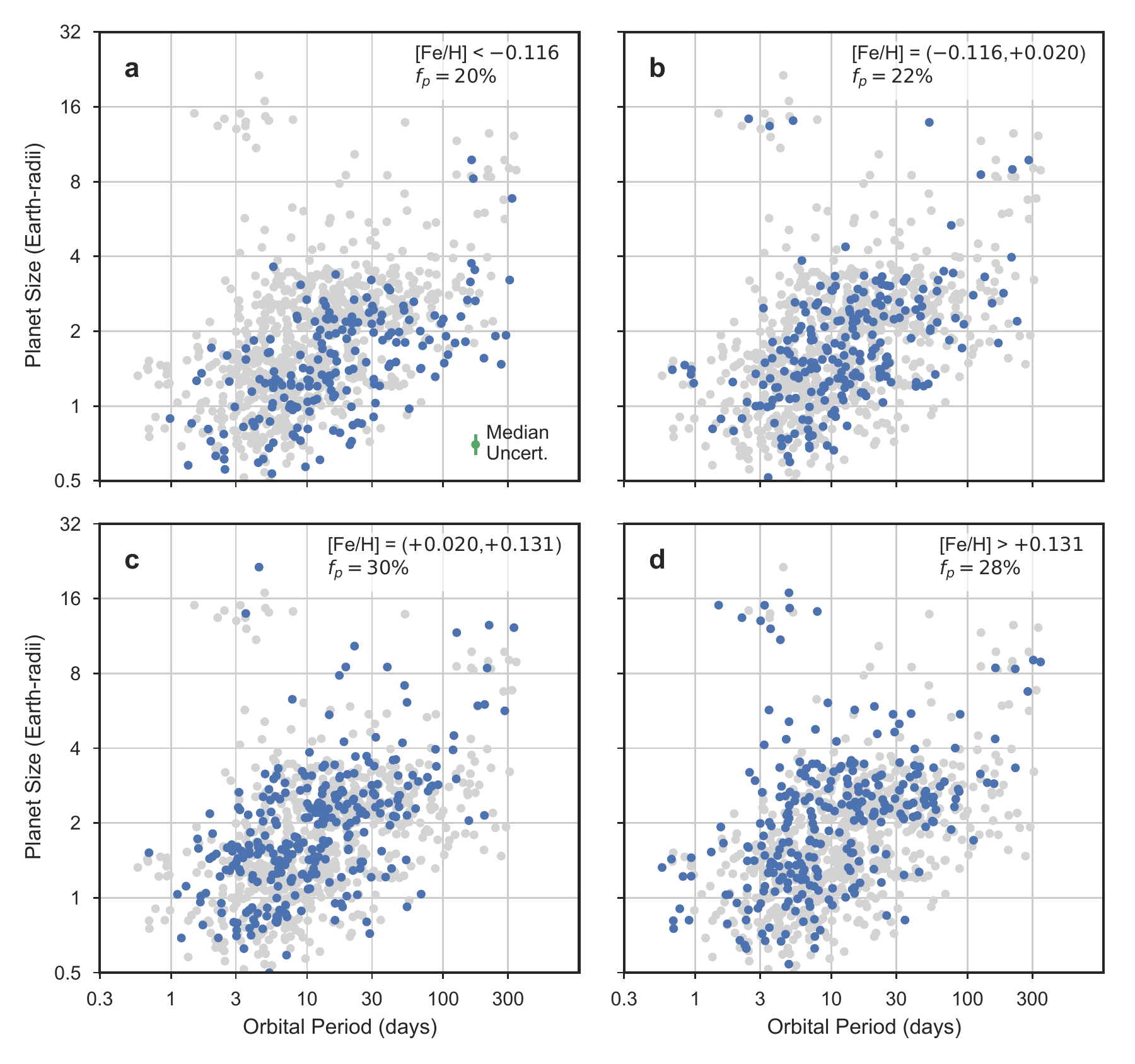}
\caption{Orbital periods and radii of planets orbiting host stars belonging to different metallicity bins. Each metallicity bin captures an equal fraction (25\%) of the parent stellar sample \ssamp (see Table~\ref{tab:smet-statistics}).
In panel (a), blue points represent planets orbiting host stars with $\fe < \samp{lamo-smet-25}$~dex, the lowest metallicity bin. The gray points show the full planet sample \psamp for context. At top right, we show $f_p$, the fraction of planets belonging to this metallicity bin. Panels (b)--(d): same as (a) except for different metallicity bins. 
If planets formed with equal efficiency regardless of host star metallicity, each bin would have $f_p$ = 25\% and the distribution of blue points would be indistinguishable from bin to bin. Planets around the lowest metallicity stars (a) are clearly confined to a restricted range of $P$--\Rp space compared to planets around the highest metallicity hosts (d), notably in the lower right envelope of longer periods and smaller sizes.\label{fig:per-prad-slices-equal-stars}}
\end{figure*}

Here, we examine the effect of host star metallicity on the 2D distribution of planet size and orbital period. We divided \psamp into four bins of host star metallicity with boundaries at $\samp{lamo-smet-25}$, $\samp{lamo-smet-50}$, and $\samp{lamo-smet-75}$~dex. The boundaries of the bins equally divide the stars in \ssamp (see Table~\ref{tab:smet-statistics}). Figure~\ref{fig:per-prad-slices-equal-stars} shows the distribution of planets in the $P$--$\Rp$ plane for the different metallicity bins. If the occurrence of planets were independent of host star metallicity, then the distribution of planets in each plot would be indistinguishable. The panels of Figure~\ref{fig:per-prad-slices-equal-stars} show clear differences, indicating that metallicity is associated with certain types of planets, and that the strength of that enhancement depends on both $P$ and \Rp.

To facilitate our investigation into the effect of metallicity across the $P$--\Rp plane, we define several regions. As a matter of convenience, we define a nomenclature for referring to these different regions. While the boundaries are matters of taste, we choose physically motivated boundaries, when possible. We consider four domains of planet size, defined below:

\begin{enumerate}
\item {\em Jupiters}. \Rp = 8--24~\Re. The lower limit is motivated by the fact that planets larger than 8~\Re tend to have masses ranging from 100 to 10,000~\Me, while planets smaller than 8~\Re tend to have lower masses ranging from 6 to 60~\Me (see, e.g., \citealt{Petigura17a}, Figure 9). At \Mp $\approx$ 100~\Me electron degeneracy pressure begins contribute significantly to a planet's pressure support \citep{Zapolsky69}. Thus 8~\Re approximates the dividing line between planets with different pressure support and interior structures. The upper radius limit of 24~\Re is based on the known sizes of hot Jupiters. The most highly irradiated Jovians discovered to date can exceed 2~\Rjup (e.g., WASP-79b, \citealt{Smalley12}).%
\footnote{Exoplanet Archive \citep{Akeson13}}%
$^{,}$%
\footnote{\url{https://exoplanetarchive.ipac.caltech.edu/}}
\item {\em Sub-Saturns}. \Rp = 4--8~\Re. Sub-Saturns are typically typically less massive the larger Jupiters. They are roughly 10$\times$ more rare than the smaller sub-Neptunes, suggesting a different formation pathway. A radius of 4~\Re approximates a breakpoint in the planet size distribution where planet occurrence rises rapidly with decreasing size (see \citealt{Fulton17}).

\item {\em Sub-Neptunes}. \Rp = 1.7--4.0~\Re. The lower radius limit corresponds to a likely transition radius between rocky planets and planets that have envelopes that are an appreciable fraction of the total planet size. Among the observations supporting such a transition are measurements of planet bulk density from transits and radial velocities \citep{Marcy14,Weiss14,Rogers15} and the observation of a gap in the radius distribution of planets by \cite{Fulton17}.

\item {\em Super-Earths}. \Rp = 1.0--1.7~\Re. Planets that are smaller than sub-Neptunes. Few planets in this size range have well-measured masses, but those that do are often consistent with rocky compositions.
\end{enumerate}

We also consider three domains of orbital period: 

\begin{enumerate}
\item {\em Hot Planets}. $P$~=~1--10~days. The upper limit of 10~days corresponds to a breakpoint in distribution of planet occurrence with orbital period \citep{Howard12,Fressin13}. Below $P \approx 10$~days, planet occurrence per $\log P$ is observed to decline, while planet occurrence is roughly constant for longer periods. 
\item {\em Warm Planets}. $P$~=~10--100~days. An intermediate range of orbital periods. While warm planets are intrinsically more common than hot planets, they represent about half of the total sample due to falling completeness and transit probability with increasing orbital period.
\item {\em Cool Planets}. $P = 100-350$~days. The longest period planets included in our survey. There are very few planets in our sample with such long periods due to falling completeness and decreasing transit probability.
\end{enumerate}

Here, we note some qualitative features in Figure~\ref{fig:per-prad-slices-equal-stars}. Stars of all metallicity bins host warm super-Earths and warm sub-Neptunes. Cool Jupiters are intrinsically rare at all metallicities, but they are present in all four metallicity bins. We observe a steady increase of hot Jupiters with increasing metallicity. Sub-Saturns of all orbital periods are almost completely absent in the lower two metallicity bins, which represent half of the parent sample \ssamp; they are almost exclusively found around high metallicity stars. While there are some examples of hot super-Earths in each metallicity bin, their numbers increase with increasing metallicity. Finally, there is almost a complete absence of hot sub-Neptunes in the lowest metallicity bins. Hot sub-Neptunes are more common with increasing metallicity.

In order to quantitatively assess the extent to which metallicity enhances the production of different types of planets, we compare the distribution of planet host metallicities to that of the field star population. In Table~\ref{tab:per-prad-stats}, we list the mean metallicities of the various planet subclasses as well as the standard error of the mean (SEM), which we compare to the mean field star metallicity, as measured from LAMOST spectra.

We assess the significance of the difference between field star and planet host star metallicities using the student $t$-test, which evaluates the difference between the means of the two samples in units of SEM, the $t$-statistic. The $t$-test also returns a \pval which is the probability that field stars and planet host stars are drawn from distributions with the same mean value.

We may observe statistically significant differences in mean metallicity for two reasons: (1) intrinsic differences between planet host and field star metallicities or (2) residual offsets in the CKS and LAMOST metallicity scales. While we calibrated LAMOST metallicities to the CKS scale, we estimate that offsets of 0.01~dex may remain (see Section~\ref{sec:sample-lamost} and Appendix~\ref{sec:calibrate-lamost}). To account for the possibility of such residual offsets, we perform three $t$-tests for each sample where we shift the LAMOST metallicities by $-0.01$~dex, 0.00~dex, and $+0.01$~dex. Each of these different tests returns a different \pval. We use the largest (most conservative) \pval to assess differences in mean metallicities. If the largest $p$-value is less than 0.01, we deem the metallicity difference to be significant.

Stars hosting Jupiter-size planets have enhanced metallicities, \femn = $\samp{All Jupiters mean}\pm\samp{Hot Jupiters sem}$~dex. The hot Jupiters hosts, with mean metallicity of \femn = $\samp{Hot Jupiters mean}\pm\samp{Hot Jupiters sem}$~dex, are significantly enhanced relative to field stars (\pval < $\samp{Hot Jupiters pval}$). While the mean metallicities for the warm and cool Jupiter hosts are also enhanced, the small numbers of such planets prevent a high significance detection of a metallicity enhancement.

Of all the planet size classes studied, the sub-Saturn hosts have the highest mean metallicity, \femn = $\samp{All Sub-Saturns mean}\pm\samp{All Sub-Saturns sem}$~dex. Among the sub-Saturns, the hot sub-Saturns have the highest mean host star metallicity, \femn = $\samp{Hot Sub-Saturns mean}\pm\samp{Hot Sub-Saturns sem}$~dex. We find that the hot and warm sub-Saturns hosts were significantly enhanced compared to field stars, while small numbers of detected cool sub-Saturns prevented a detailed comparison.

As a whole, the sub-Neptunes have a mean metallicity of \femn = $\samp{All Sub-Neptunes mean}\pm\samp{All Sub-Neptunes sem}$~dex, close to the field star value. However, when split according to orbital period, we find that the  hot sub-Neptunes have an enhanced mean metallicity, \femn = $\samp{Hot Sub-Neptunes mean}\pm\samp{Hot Sub-Neptunes sem}$~dex. The large mean metallicity, combined with the large number of such planets ($N = \samp{Hot Sub-Neptunes n}$), results in a very significant detection of a metallicity enhancement (\pval < $\samp{Hot Sub-Neptunes pval}$). Due to their high detectability and high intrinsic occurrence, the warm sub-Neptunes have the largest total number of any $P$--\Rp subclass, $N = \samp{Warm Sub-Neptunes n}$. The large number of planets means that even small offsets in mean metallicities can be significant. While the mean metallicity of warm sub-Neptunes, \femn = $\samp{Warm Sub-Neptunes mean}\pm\samp{Warm Sub-Neptunes sem}$~dex), was not as high as that of the hot sub-Neptunes, it was significantly elevated relative to field stars (\pval < $\samp{Warm Sub-Neptunes pval}$).

Finally, the super-Earth-size planets have the lowest mean metallicity of $\samp{All Super-Earths mean}\pm\samp{All Super-Earths sem}$~dex, which is consistent with field star metallicity. However, as with the sub-Neptunes, the hot super-Earth hosts exhibit enhanced metallicity of $\samp{Hot Super-Earths mean}\pm\samp{Hot Super-Earths sem}$~dex. The difference in mean metallicity (in dex) is not as large for the sub-Neptunes, so while the offset is still significant (\pval < $\samp{Hot Super-Earths pval}$), it is not as significant as for the hot sub-Neptunes. Despite the large number of warm super-Earths, we cannot detect a significant offset in mean metallicity.

One may wonder whether the mean metallicities of stars hosting warm super-Earths/sub-Neptunes could possibly be different from the mean metallicity of the \Kepler field ($-$0.01~dex), given the high intrinsic occurrence of these planets. While numerous previous works have shown that warm super-Earths and sub-Neptunes are intrinsically common (see, e.g., \citealt{Petigura13b}), neither class of planets is found around 100\% of stars. For example, in Section~\ref{sec:occur-per-radius} we show that there are $\approx$17 warm super-Earths per 100 stars. Therefore, the mean metallicity of warm super-Earth hosts could be significantly different than that of field stars. We consider the following limiting case: Suppose that the process that produces warm super-Earths has a step function dependence on metallicity. Among \Kepler targets, 17\% of stars have \fe > 0.16 dex. A universe where every star with \fe > 0.16~dex produced one warm super-Earth, and every star with \fe < 0.16~dex produced zero, would be consistent with the measured occurrence. By consulting the distribution of \Kepler field star metallicities (see Section~\ref{sec:sample-lamost}), we find that the average metallicity of all stars with \fe > 0.16~dex is 0.25~dex. In this limiting case, the average metallicity of warm super-Earth hosts would be \femn = 0.25~dex, much higher than the observed value of $\samp{Warm Super-Earths mean}\pm\samp{Warm Super-Earths sem}$~dex. That warm super-Earth and sub-Neptune hosts have mean metallicities nearly the same as that of field stars demonstrates qualitatively that their occurrence is not a steep function of metallicity. We show this quantitatively in Section~\ref{sec:occur-per-radius-smet}.

In summary, close-in ($P < 10$~days) planets of all sizes have enhanced metallicity host stars. Warm super-Earth-size planets at intermediate distances ($P$~=~10--100~days) orbit stars with a metallicity distribution similar to that of field stars. Larger planets in this period range have enhanced metallicities, and the sub-Neptunes and sub-Saturns have a significant metallicity enhancement. Finally, the number of detected cool planets ($P$~=~100--350~days) of all sizes is too small to permit the detection of different host star metallicities.

\begin{deluxetable*}{lrrrRRRr}
\tablecaption{Comparison of Planet Host and Field Star Metallicities\label{tab:per-prad-stats}}
\tablecolumns{8}
\tablehead{
\colhead{Name} & 
\colhead{$P$} & 
\colhead{$\Rp$} & 
\colhead{$n_\star$} &
\colhead{\femn} &
\colhead{$t$-stat} &
\colhead{$p$-value} &
\colhead{Sig\tablenotemark{a}} \\
\colhead{} & 
\colhead{day} & 
\colhead{\Re} & 
\colhead{} &
\colhead{dex} &
\colhead{} &
\colhead{} &
\colhead{}
}
\tablewidth{0pt}
\startdata
Hot Jupiters & 1--10 & 8.0--24.0 & 14 & +0.19 \pm 0.04 & 4.5 & < \num{8e-04} & Y \\
Warm Jupiters & 10--100 & 8.0--24.0 & 4 & +0.06 \pm 0.02 & 3.4 & < \num{6e-02} & N \\
Cool Jupiters & 100--350 & 8.0--24.0 & 13 & +0.06 \pm 0.05 & 1.4 & < \num{3e-01} & N \\
Hot Sub-Saturns & 1--10 & 4.0--8.0 & 7 & +0.26 \pm 0.04 & 6.5 & < \num{7e-04} & Y \\
Warm Sub-Saturns & 10--100 & 4.0--8.0 & 19 & +0.14 \pm 0.02 & 5.6 & < \num{6e-05} & Y \\
Cool Sub-Saturns & 100--350 & 4.0--8.0 & 7 & +0.11 \pm 0.06 & 1.9 & < \num{1e-01} & N \\
Hot Sub-Neptunes & 1--10 & 1.7--4.0 & 108 & +0.11 \pm 0.02 & 7.0 & < \num{4e-09} & Y \\
Warm Sub-Neptunes & 10--100 & 1.7--4.0 & 282 & +0.04 \pm 0.01 & 4.0 & < \num{2e-03} & Y \\
Cool Sub-Neptunes & 100--350 & 1.7--4.0 & 29 & -0.05 \pm 0.03 & -1.3 & < \num{3e-01} & N \\
Hot Super-Earths & 1--10 & 1.0--1.7 & 181 & +0.05 \pm 0.01 & 4.7 & < \num{2e-04} & Y \\
Warm Super-Earths & 10--100 & 1.0--1.7 & 132 & -0.04 \pm 0.02 & -2.2 & < \num{1e-01} & N \\
Cool Super-Earths & 100--350 & 1.0--1.7 & 4 & -0.36 \pm 0.03 & -11.4 & < \num{2e-03} & Y \\
All Jupiters & 1--350 & 8.0--24.0 & 31 & +0.12 \pm 0.03 & 4.2 & < \num{6e-04} & Y \\
All Sub-Saturns & 1--350 & 4.0--8.0 & 33 & +0.16 \pm 0.02 & 7.2 & < \num{1e-07} & Y \\
All Sub-Neptunes & 1--350 & 1.7--4.0 & 419 & +0.05 \pm 0.01 & 6.2 & < \num{7e-07} & Y \\
All Super-Earths & 1--350 & 1.0--1.7 & 317 & +0.01 \pm 0.01 & 1.3 & < \num{8e-01} & N \\
\enddata
\tablecomments{We inspected the metallicity distribution of various groups of planets defined by their sizes and orbital periods. For each group, we computed the mean metallicity \femn and the standard error of the mean. We compare these metallicities to the metallicities of \Kepler field stars, as measured by LAMOST, using the student $t$-test. This test returns a $t$-statistic, which quantifies the statistical difference in mean metallicities, and a $p$-value, the probability that the two samples were drawn from distributions with the same mean. }
\tablenotetext{a}{Is the difference between planet host and field star metallicities significant (i.e., is \pval < $10^{-2}$)?}
\end{deluxetable*}

\section{Planet occurrence: Methodology}
\label{sec:occur-methodology}
In Section~\ref{sec:smet}, we examined the number distribution of planets belonging to host stars of differing metallicities. We now address the underlying prevalence of planets. In this section, we our methodology for computing planet occurrence and fitting parameterized descriptions of planet occurrence to the observed data. Section~\ref{sec:occur-per-radius} presents planet occurrence as a function of orbital period and planet size. Section~\ref{sec:occur-per-radius-smet} presents planet occurrence as a function of orbital period, planet size, and stellar metallicity.

\subsection{Definitions}
\label{sec:methodology-definitions}
Adopting the notation of \cite{Youdin11}, the probability that a star with properties $\pmb{z}$ has a planet with properties $\pmb{x}$ that lie in a volume of $d \pmb{x}$ of phase space is
\begin{equation}
d f = \frac{\partial f(\pmb{x},\pmb{z})}{\partial {\pmb{x}}} d \pmb{x}.
\end{equation}
In this paper, $\pmb{z}$ is metallicity $M$ = \fe%
\footnote{We use $M$ interchangeably with \fe for compactness.}
and $\pmb{x}$ is some combination of $\log P$ and $\log \Rp$.%
\footnote{In this paper, $\log$ only refers to $\log_{10}$. Natural logs are always expressed as $\ln$.}
As a matter of convenience we express differential distribution as 
\begin{equation}
\frac{\partial f(\pmb{x},\pmb{z})}{\partial {\pmb{x}}} = C g (\pmb{x},\pmb{z}),
\end{equation}
where $C$ is a normalization constant and $g$ is a shape function that may depend on stellar and/or planetary properties.

The number of planets per star (NPPS) having stellar properties $\pmb{z}$ over a specified volume of planet properties $X$ is
\begin{equation}
f(\pmb{z}) = \int_{X} C g (\pmb{x},\pmb{z}) d\pmb{x}
\end{equation}
The NPPS having stellar properties within a range of stellar properties $Z$ is
\begin{equation}
f = \frac{1}{n_{\star}} \sum_{j} \int_{X} C g (\pmb{x},\pmb{z}_j) d\pmb{x}
\end{equation}
where $j$ labels stars where $z_j \in Z$.

Note that our definition of planet occurrence is NPPS. As a matter of convenience, we often express this in units of planets per 100 stars. Some papers (e.g. \citealt{Fischer05}) define planet occurrence as the fraction of stars with planets (FSWP). For a transit survey like \Kepler, computing FSWP is challenging because one must compute the number of stars {\em without planets}. However, a transit survey cannot distinguish between stars {\em without planets} and stars {\em without transiting planets}.  Converting between NPPS and FSWP requires detailed modeling of planet multiplicity and mutual inclinations and is beyond the scope of this work. For more discussion on these points, see \cite{Youdin11} and references therein.

\subsection{Occurrence within a Cell}
\label{sec:methodology-cell}

In this paper, we compute and display planet occurrence over rectilinear ``cells'' constructed from various combinations of $\log P$, $\log \Rp$, and $M$. We express the size of these cells as $\Delta \log P$, $\Delta \log \Rp$, $\Delta M$, measured in dex. The occurrence within a cell is \fcell, which depends on the number of independent trials \ntrial that could have yielded a detected planet. In the limit where every star in \ssamp has one transiting planet that is also detectable by the \Kepler pipeline,  \ntrial~=~\nstar. In practice, \ntrial must account for (1) non-transiting orbital inclinations and (2) lack of detectability due to insufficient S/N. For a given planet size and orbital period,
\begin{equation}
\ntrial = \sum_{i=1}^{\nstar}\, \ptrr{i}(P)\, \pdett{i}(P, \Rp)
\end{equation}
where $i$ labels each star in \ssamp, \ptrr{i} is the transit probability, and \pdett{i} is the probability that a transiting planet would be detectable by the \Kepler pipeline. We represent this more compactly as
\begin{equation}
\ntrial = \nstar \left< \ptrr{i} \, \pdett{i} \right>,
\end{equation}
where $\left< \cdot \right>$ denotes the arithmetic mean.

Following \cite{Bowler15}, if we assume that planet occurrence is log-uniform within a cell, $\ntrial$ for a cell of size $\Delta \pmb{x} = \Delta \log P \times \Delta \log \Rp$ is
\begin{eqnarray}
\label{eqn:ntrials-per-prad}
\ntrial & = & \frac{\nstar}{\Delta \pmb{x}} \int \left<\ptr\pdet \right> d \pmb{x}.
\end{eqnarray}
For a given cell with \ntrial trials, \npl detected planets, and \nnd = $\ntrial - \npl$ non-detections, the likelihood of \fcell is given by the binomial distribution%
\begin{eqnarray}
P (\fcell|\npl,\ntrial) & = & P (\npl|\fcell,\ntrial) \\
                            & = & C \fcell^{\npl} (1 - \fcell)^{\nnd} \label{eqn:likelihood}
\end{eqnarray}
where $C$ is the following normalization constant:%
\footnote{Here, the factorials have been replaced with gamma functions via $\Gamma(x+1) = x!$ and $(n+1)$ is a normalization constant.}
\begin{equation}
C = \frac{(\ntrial + 1) \Gamma(\ntrial + 1)}{\Gamma(\npl+1) \Gamma(\nnd + 1)}.
\end{equation}
If no transits are detected in a given cell, we may place an upper bound on the occurrence rate by calculating the maximum value of $\fcell$ that yields zero planet detections ($\npl = 0$) with 90\% probability. This is done by numerically finding \fcell that satisfies
\begin{equation}
\int_{0}^{\fcell} P (f|\npl, 0) df = 90\%.
\end{equation}
When analyzing planet occurrence as a function of metallicity, we must consider the fraction of stars in \ssamp that falls within a specified range of metallicity. Therefore, to compute \fcell for a cell bounded by $[\Rpp{1}, \Rpp{2}]$, $[\PP{1}, \PP{2}]$, and $[\smet{1}, \smet{2}]$ we simply multiply \ntrial from Equation~\ref{eqn:ntrials-per-prad} by the fraction of LAMOST stars with metallicities between \smet{1} and \smet{2}. This treatment assumes that planet detectability does not depend on metallicity. We justify this assumption in Appendix~\ref{sec:smet-detectability}.

\subsection{Completeness}
\label{sec:methodology-completeness}
To compute occurrence, we must estimate $\left<\ptr\pdet\right>$. The probability that a randomly inclined planet at a distance of $a$ will transit with $b < \val{b-cut}$ is $\ptr = \val{b-cut} \Rstar / a$, assuming circular orbits. We compute $\ptr$ for each star using the tabulated values of \Rstar and \Mstar from the Q1--Q17 (DR25) stellar properties table \citep{Mathur17}, along with Kepler's Third Law. 

The probability of detecting a transiting planet with size \Rp and period $P$ depends on number of factors, the most significant of which is the S/N. We follow the methodology of \cite{Fulton17} and compute an expected S/N using the following formula:
\begin{equation}
\mathrm{S/N} = 
\left(\frac{\Rp}{\Rstar}\right)^{2}
\sqrt{\frac{T_{obs}}{P}}
\left(\frac{1}{\sigma(T_{14})}\right)
\end{equation}
Here, $T_{\mathrm{obs}}$ is the time the star was observed by \Kepler and $\sigma(T_{14})$ represents the photometric variability on transit timescales. The DR25 stellar properties table lists $T_{\mathrm{obs}}$, \Rstar, and photometric noise%
\footnote{Combined Differential Photometric Precision (CDPP, \citealt{Jenkins10})}
on 3, 6, and 12~hr timescales. We compute $\sigma(T_{14})$ at intermediate values of $T_{14}$ through piecewise linear interpolation (or extrapolation) in $\log \sigma$--$\log T_{14}$ space.

\begin{figure*}
\centering
\includegraphics{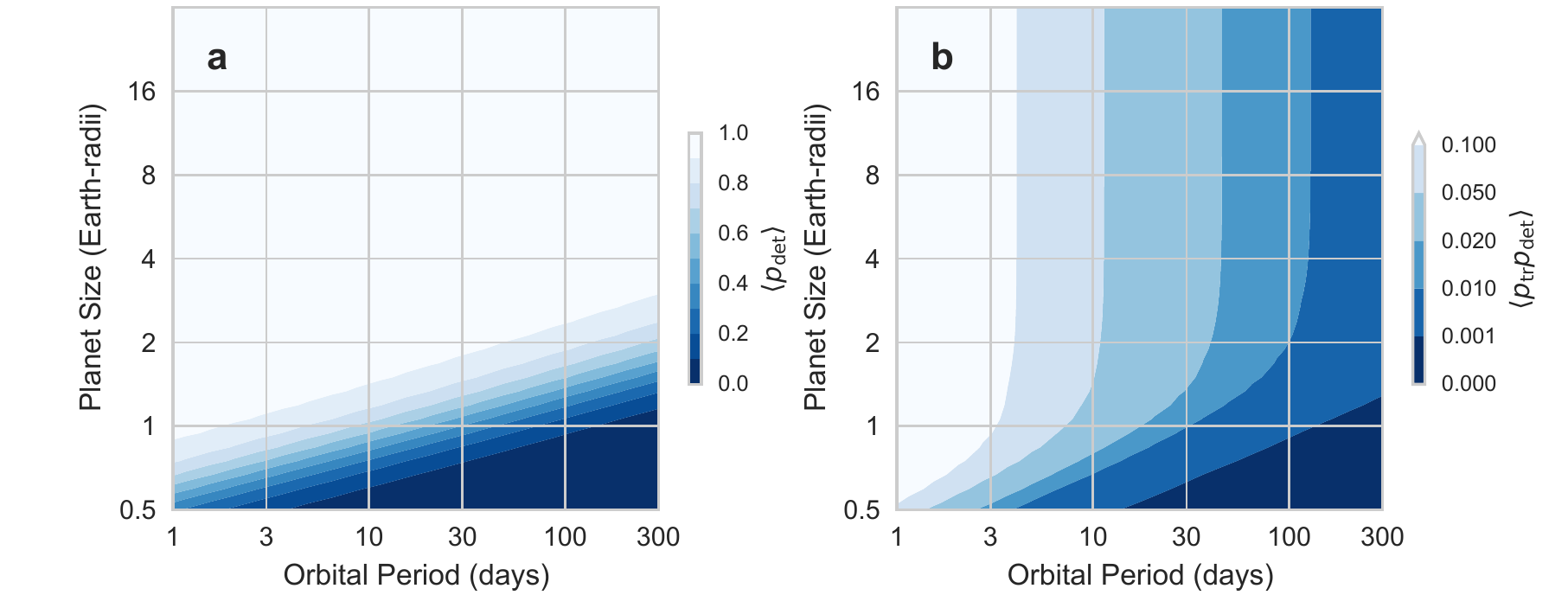}
\caption{Average detectability of planets as a function of planet size and orbital period. Panel (a): the probability that a transiting planet would be detected averaged overall stars in our sample, $\left<p_{det}\right>$. Panel (b): the product of the transit and detection probability averaged overall stars in the sample $\left<p_{det} p_{tr}\right>$. The effective number of stars from which the planet detections are drawn is simply $\ntrial = \nstar \left<p_{det} p_{tr}\right>$.}
\vspace{0.1in}
\label{fig:prob-detect+transit}
\end{figure*}

Characterizing \pdet as a function of S/N has been addressed several previous works (e.g. \citealt{Fressin13,Petigura13b,Christiansen15}). This is a challenging problem, which depends on how a complex, multi-stage transit pipeline performs in the face of correlated, non-stationary, and non-Gaussian photometric noise. We adopted the \pdet(S/N) of \cite{Fulton17}, who used the transit injections of \cite{Christiansen15} to derive the following relationship:
\begin{equation}
\pdet(s) = \Gamma(k) \int_0^{\frac{s - l}{\theta}} t^{k-1} e^{-t} dt 
\end{equation}
where $s$ is the S/N, $k = 17.56$, $l = 1.00$, and $\theta = 0.49$. We illustrate $\left<\pdet\right>$ and $\left<\ptr\pdet\right>$ as a function of $P$ and $\Rp$ in Figure~\ref{fig:prob-detect+transit}.

\subsection{Fitting the Occurrence Distribution}
\label{sec:methodology-fitting}
We often wish to characterize the occurrence distribution with parametric models. These models serve to quantify shapes and trends in the population of planets. We extend the maximum-likelihood method of \cite{Howard12} to fit $C g(\pmb{x},\pmb{z}; \pmb{\theta})$ to the observed planet population, where $\pmb{\theta}$ represents the vector of shape parameters in our model.

The occurrence within each cell \fcell gives an estimate of the differential occurrence rate via
\begin{equation}
C g(\pmb{x},\pmb{z}; \pmb{\theta}) = \frac{\fcell}{\Delta \pmb{x}}
\end{equation}
The log-likelihood of a given model with parameters $\left\{C,\pmb{\theta}\right\}$ given the observed rate in a single cell labeled by $i$ is
\begin{equation}
\label{eqn:loglike-cell}
\ln L_{i} = n_{\mathrm{pl},i} \ln C g \Delta \pmb{x}+ n_{\mathrm{nd},i} \ln (1 - C g\Delta \pmb{x}).
\end{equation}
Each cell is an independent constraint on $C g(\pmb{x},\pmb{z}; \pmb{\theta})$, so a fit to multiple cells requires maximizing the combined log-likelihood over all cells:
\begin{equation}
\ln L =  \sum_{i=1}^{\ncell} L_{i}.
\end{equation}

When fitting parameterized models of occurrence in Sections~\ref{sec:occur-per-radius} and \ref{sec:occur-per-radius-smet}, we use very small bins having $\Delta \log P$ = 0.05~dex and $\Delta \log M$ = 0.05~dex. As a result, many bins have one or zero planet detections. The maximum-likelihood method is stable for small bin sizes because Equation~\ref{eqn:loglike-cell} incorporates non-detections. We experimented with even finer bins, but did not observe significant changes in the results. After finding the max-likelihood model, we explore the credible range of models using Markov Chain Monte Carlo (MCMC) sampling.%
\footnote{We used the affine invariant sampler of \cite{Goodman10} as implemented in Python by \cite{Foreman-Mackey13}.}

\section{Planet Occurrence: Period and Radius}
\label{sec:occur-per-radius}
\begin{figure*}
\centering
\includegraphics{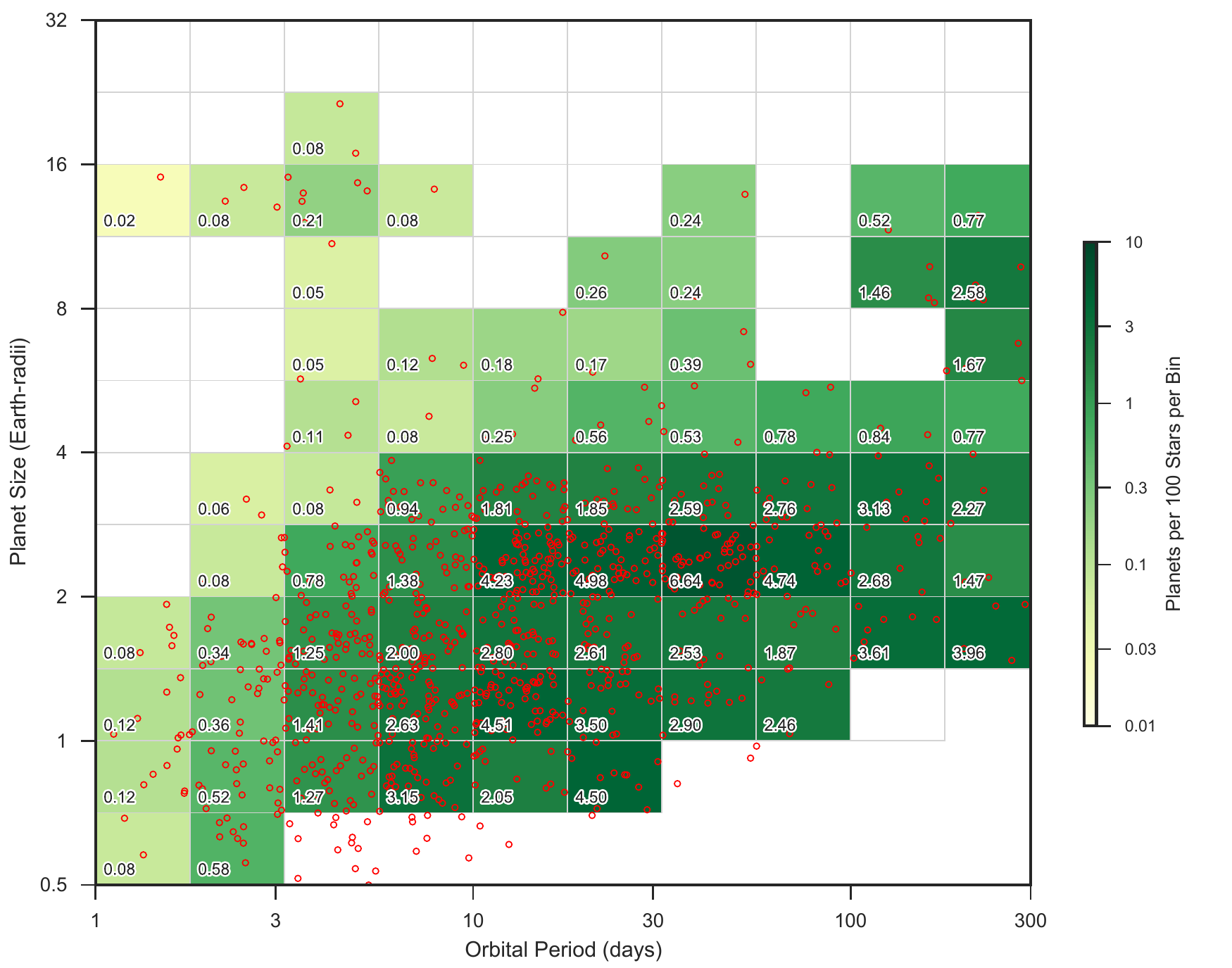}
\caption{The domain of orbital period and planet size, which has been sub-divided into numerous sub-domains, ``cells.'' Each cell is annotated with the average number of planets per 100 stars having properties within each cell. This quantity is also reflected through the shading of each cell. The number of planets has been corrected for the probability of transiting and for detection completeness.\label{fig:checkerboard}}
\end{figure*}

We first present planet occurrence on the $P$--\Rp plane, averaging over stellar metallicity. Following \cite{Howard12}, we divided the domain of orbital period and planet size into numerous cells spanning $\Delta \log P \times \Delta \log \Rp$ =  0.25~dex $\times$ 0.15~dex. For each cell, we computed \fcell according to the prescription from Section~\ref{sec:methodology-cell}. We display \fcell as a checkerboard in Figure~\ref{fig:checkerboard}, where each cell in is shaded and annotated according to the occurrence of planets within the cell. We list the cell-by-cell rates, uncertainties, and upper limits in Table~\ref{tab:occurrence} in the Appendix. 

\begin{figure*}
\centering
\includegraphics[width=0.7\textwidth]{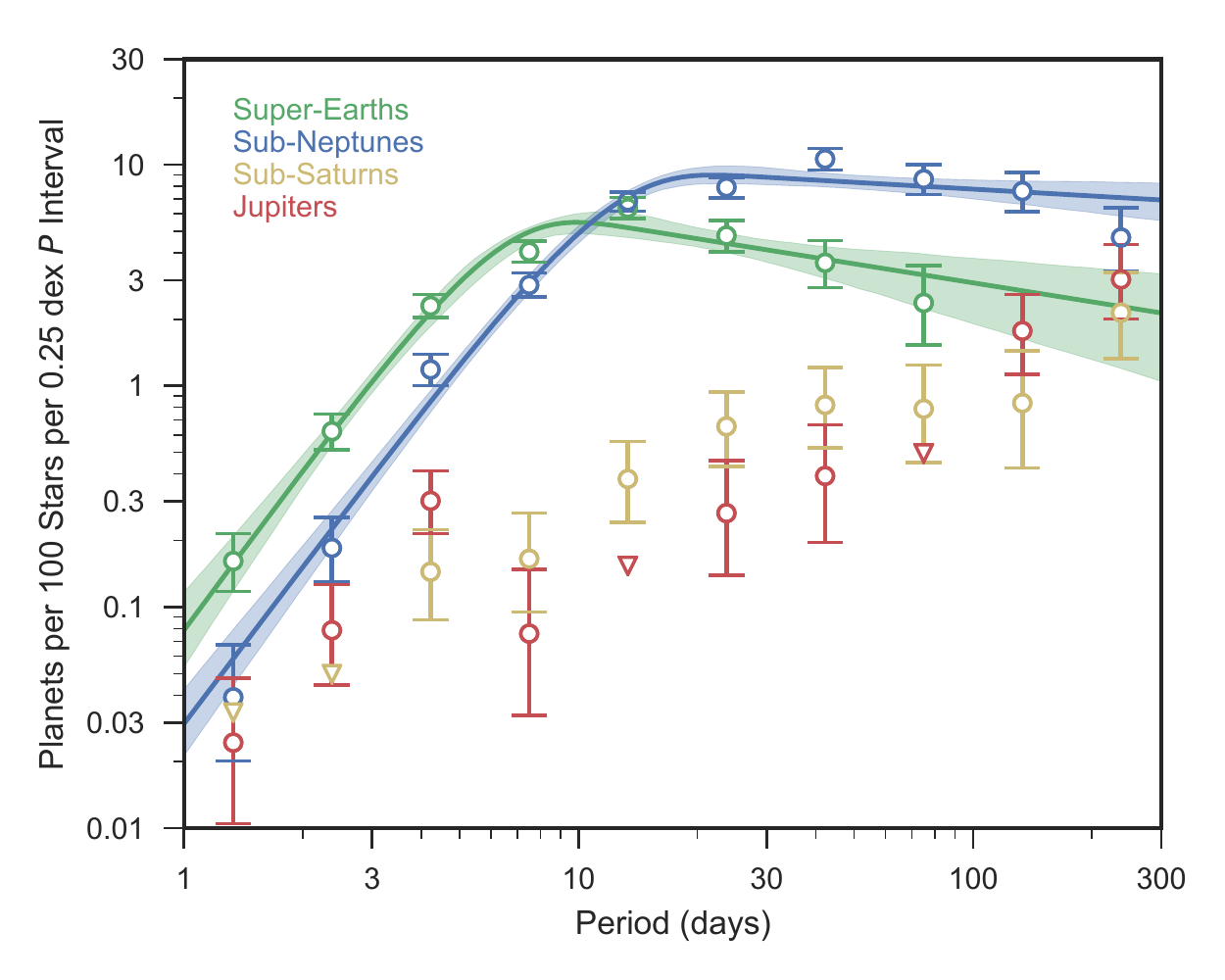}
\caption{Planet occurrence as a function of orbital period for various size classes. For example, green points show the number of super-Earths per 100 stars per 0.25~dex interval in period. Downward arrows represent upper limits (90\%). For the super-Earths and sub-Neptunes, we fit the occurrence using the power law and exponential cutoff model described in Section~\ref{sec:occur-per-radius} (Equation~\ref{eqn:powerlaw-cutoff}). The solid lines and bands show the best-fitting model and 1$\sigma$ range of credible models, respectively. For super-Earths and sub-Neptunes at $P < 10$~days, planet occurrence increases like $d f/d \log P \propto P^{\alpha}$, where $\alpha \approx$ \fit{per-all-se-gamma+beta} and $\alpha \approx$ \fit{per-all-sn-gamma+beta}, respectively. At longer orbital periods, occurrence is nearly uniform in $ \log P$. A transition period $P_0$ characterizes where the distributions changes slope, which occurs at $P_0$ =  \fit{per-all-se-per0}~days and $P_0$ = \fit{per-all-sn-per0}~days, respectively. The distributions of sub-Saturns and Jupiters are not well-described by this model. Their occurrence gradually increases over $P$ = 1--300~days.\label{fig:occur-per}}
\end{figure*}

We also show occurrence as a function of orbital period for various size classes in Figure~\ref{fig:occur-per}, computed over bins spanning $\Delta \log P$ = 0.25~dex. Errorbars and upper limits are computed according to the methodology presented in Section~\ref{sec:methodology-cell}. For super-Earths and sub-Neptunes, inspection of the binned occurrence rates reveals that \dfdlogp increases with $P$ until a transition period $P_0$, above which occurrence is nearly uniform in $\log P$. This ``knee'' points toward an important orbital distance in the formation or subsequent migration of planets. We characterized the period distribution with the following parameterization from \cite{Howard12}:
\begin{equation}
\label{eqn:powerlaw-cutoff}
\frac{d f}{d \log P} = C P^{\beta} \left(1 - e^{-(P/P_0)^{\gamma}} \right).
\end{equation}
Far from $P_0$, this function reverts to a power law with the following indices:
\begin{eqnarray}
\frac{d f}{d \log P} \propto 
  \begin{cases}
     P^{\alpha}          &  \text{if $P \ll P_0$, where $\alpha = \gamma + \beta$} \\
     P^{\beta}           &  \text{if $P \gg P_0$} 
  \end{cases}
\end{eqnarray}
We fit this distribution using the methodology from Section~\ref{sec:methodology-cell} and list the associated parameters in Table~\ref{tab:fit-per}. Our best-fit model and 1$\sigma$ range of credible models are shown in Figure~\ref{fig:occur-per}. As explained in Section~\ref{sec:methodology-fitting}, the max-likelihood fitting uses much finer bins than those displayed in Figure~\ref{fig:occur-per}. The binned rates shown in Figure~\ref{fig:occur-per} serve to guide the eye and are not used in the fitting.

For super-Earths the transition period $P_0$ is \fit{per-all-se-per0}~days. At shorter orbital periods, occurrence rises rapidly with increasing $P$, with $df \propto P^{\alpha} d \log P$, where $\alpha$ = \fit{per-all-se-gamma+beta}. At longer orbital periods, $df \propto P^{\beta} d \log P$, where $\beta$ = \fit{per-all-se-beta}. The transition period for sub-Neptunes is farther out at $P_0$ = \fit{per-all-sn-per0}~days, but the power law indices are similar, with $\alpha$ = \fit{per-all-sn-gamma+beta} and $\beta$ = \fit{per-all-sn-beta}. The distributions of sub-Saturns and Jupiters are not well-described by the power law cutoff model. Their occurrence gradually increases over $P$ = 1--300~days.

\begin{figure*}
\includegraphics[width=1.0\textwidth]{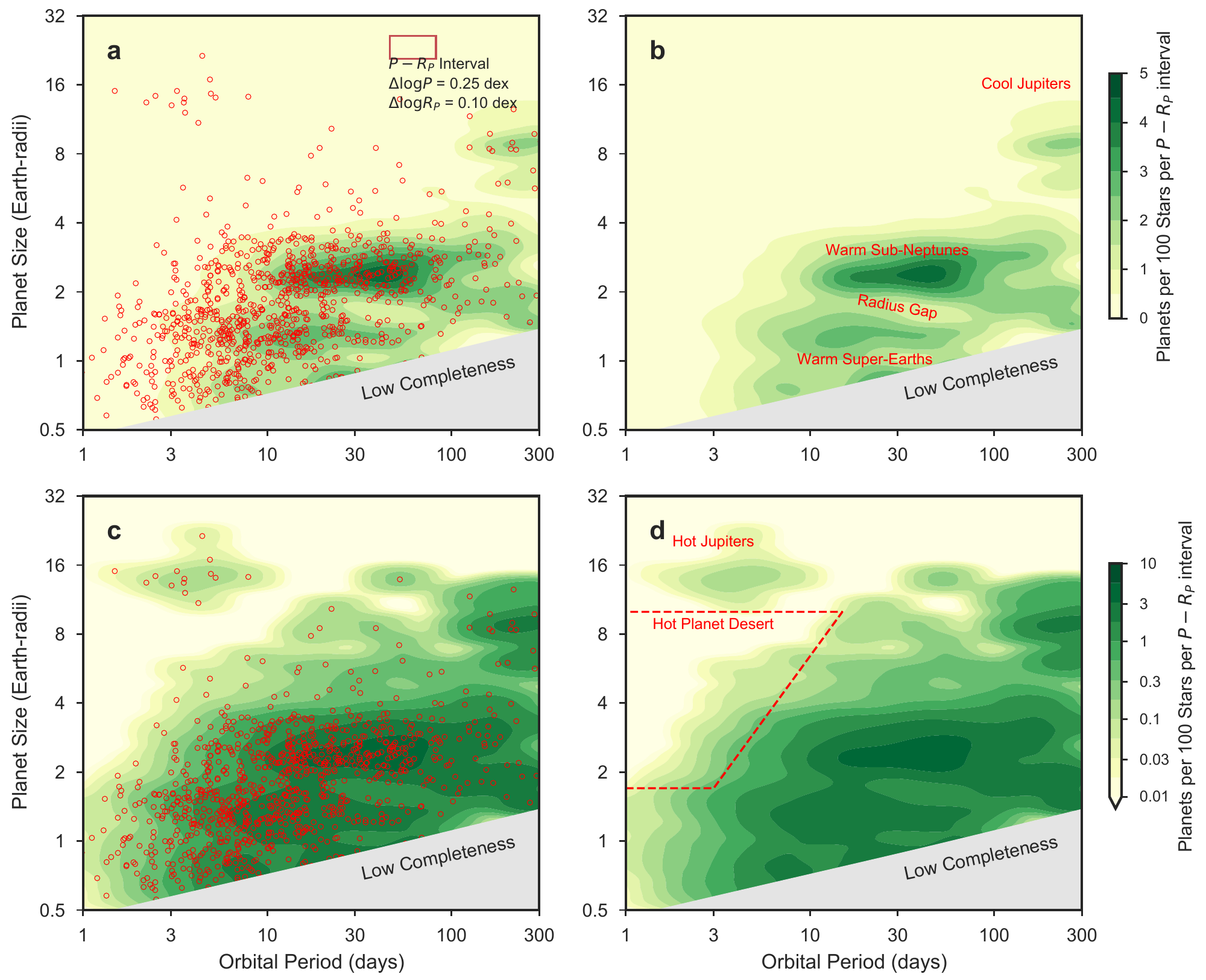}
\caption{Panel (a) shows the planet sample \psamp and planet occurrence in the $P$--\Rp plane. The shading at each ($P$,\Rp) point represents the number of planets per 100 stars within an interval centered at ($P$,\Rp) that spans 0.25~dex in $\log P$ and 0.10~dex in $\log \Rp$. The size of the interval is indicated at top right.
For example, the darkest contour indicates that there are $\approx$4 planets per 100 Sun-like stars having periods within 0.125~dex  of 40~days and radii within 0.05~dex of 2.5~\Re.
Panel (b): same as (a), but without detected planets for clarity. Warm sub-Neptunes are the most abundant class of planets, and warm super-Earths are another common class of planets. The super-Earths and sub-Neptunes are separated by a diagonal gap of low planet occurrence, described by \cite{Fulton17}. 
Panels (c)--(d): same as (a)--(b), but with logarithmic shading to highlight domains of low occurrence. The hot Jupiter population is an ``island,'' distinct from the rest of the planet population. There is a ``desert'' of hot planets having intermediate sizes.\label{fig:contour}}
\end{figure*}

Figures~\ref{fig:checkerboard} and \ref{fig:occur-per} are convenient tools for calculating planet occurrence over various domains of period and radius. However, several features in the planet population are hard to see in these plots due to the coarse bin sizes and the arbitrary location of the bin boundaries. Figure~\ref{fig:contour} shows a finer view of planet occurrence. We computed occurrence in bins of period and radius spanning 0.25~dex and 0.10~dex respectively. We then shifted the bins in small steps of $\Delta \log P$ and $\Delta \log \Rp$ to smoothly trace out planet occurrence in the $P$--\Rp plane. The shading in Figure~\ref{fig:contour} gives a bird's eye view of the prevalence of various types of planets.

Panels (a) and (b) of Figure~\ref{fig:contour} show occurrence on a linear scale to highlight the most abundant types of planets (super-Earths and sub-Neptunes). We note the rapid increase in occurrence for $\Rp \lesssim 4~\Re$, which has been noted by \cite{Howard12} and numerous subsequent works. Recently, \cite{Fulton17} re-examined the radius distribution of small planets using the CKS catalog. \cite{Fulton17} also noted a gap in the radius distribution at \Rp~=~1.7~\Re. This gap was predicted by various groups who modeled the erosion erosion of envelopes (e.g. \citealt{Lopez13,Owen13,Jin14,Chen16}). We also resolve the gap that separates the super-Earth and sup-Neptune populations.

Panels (c) and (d) of Figure~\ref{fig:contour} show occurrence on a logarithmic scale to highlight domains of low planet occurrence. Although hot Jupiters are rare, they constitute a distinct island in the $P$--\Rp plane, surrounded by a sea of still lower occurrence. Hot planets of intermediate sizes are very rare. This triangular ``Hot Planet Desert'' has been noted by \cite{Mazeh16} and \cite{Dong17}, and is thought to be due to photo-evaporative envelope stripping. 

 We also observe an factor of $\approx$5--10~increase in occurrence of Jupiters from $P$ = 100--300~days. \cite{Cumming08} analyzed the planets detected by radial velocities from the Keck Planet Search and observed that giant planet occurrence increases by a factor $\approx$5 from $P$ = 100--300~days. This rise in the occurrence of Jovian planets between 100--300~days, observed in both the \Kepler stellar population and among nearby stars, may be associated with an ice-line at $\sim$1~au.
 
The joint distribution of planets in the $P$--\Rp plane shown in Figure~\ref{fig:contour} is an important quantitative description of the population of planets. For example, this distribution is useful for studies predicting planet yields in future surveys (e.g. \citealt{Sullivan15}). To facilitate such studies, we sample this distribution and provide the periods and radii for a representative population of planets in a sample of 100,000 Sun-like stars in Table~\ref{tab:population}. This synthetic population also offers an convenient way to compute integrated planet occurrence (but not uncertainties) over arbitrary bins of $P$ and $\Rp$.

\begin{deluxetable*}{lllllll}
\tablecaption{Best-fit Parameters for Planet Period Distributions\label{tab:fit-per}}
\tablecolumns{7}
\tablehead{
\colhead{Size Class} & 
\colhead{$\Rp$} & 
\colhead{\fe} & 
\colhead{$\log_{10} C $} & 
\colhead{$\beta$} &
\colhead{$P_0$} &
\colhead{$\gamma$} 
}
\tablewidth{0pt}
\startdata
Super-Earth & 1.0--1.7 & all & \fit{per-all-se-logkp} & \fit{per-all-se-beta} & \fit{per-all-se-per0} & \fit{per-all-se-gamma} \\ 
            &          & < 0 & \fit{per-sub-se-logkp} & \fit{per-sub-se-beta} & \fit{per-sub-se-per0} & \fit{per-sub-se-gamma} \\ 
            &          & > 0 & \fit{per-sup-se-logkp} & \fit{per-sup-se-beta} & \fit{per-sup-se-per0} & \fit{per-sup-se-gamma} \\ 
Sub-Neptune & 1.7--4.0 & all & \fit{per-all-sn-logkp} & \fit{per-all-sn-beta} & \fit{per-all-sn-per0} & \fit{per-all-sn-gamma} \\ 
            &          & < 0 & \fit{per-sub-sn-logkp} & \fit{per-sub-sn-beta} & \fit{per-sub-sn-per0} & \fit{per-sub-sn-gamma} \\ 
            &          & > 0 & \fit{per-sup-sn-logkp} & \fit{per-sup-sn-beta} & \fit{per-sup-sn-per0} & \fit{per-sup-sn-gamma} \\ 
\enddata
\tablecomments{Best-fit parameters associated with the occurrence model defined in Equation~\ref{eqn:powerlaw-cutoff}.}
\end{deluxetable*}

\section{Planet Occurrence: Period, Radius, and Stellar Metallicity}
\label{sec:occur-per-radius-smet}
Here, we analyze planet occurrence as a function of period, radius, and stellar metallicity. As a first step, we divided the super-Earth and sub-Neptune populations into two groups, depending on whether their host stars had sub- or super-solar metallicities. We then repeated the analysis from Section~\ref{sec:occur-per-radius}, modeling the occurrence distribution of both groups according to Equation~\ref{eqn:powerlaw-cutoff}.

\begin{figure*}
\centering
\includegraphics{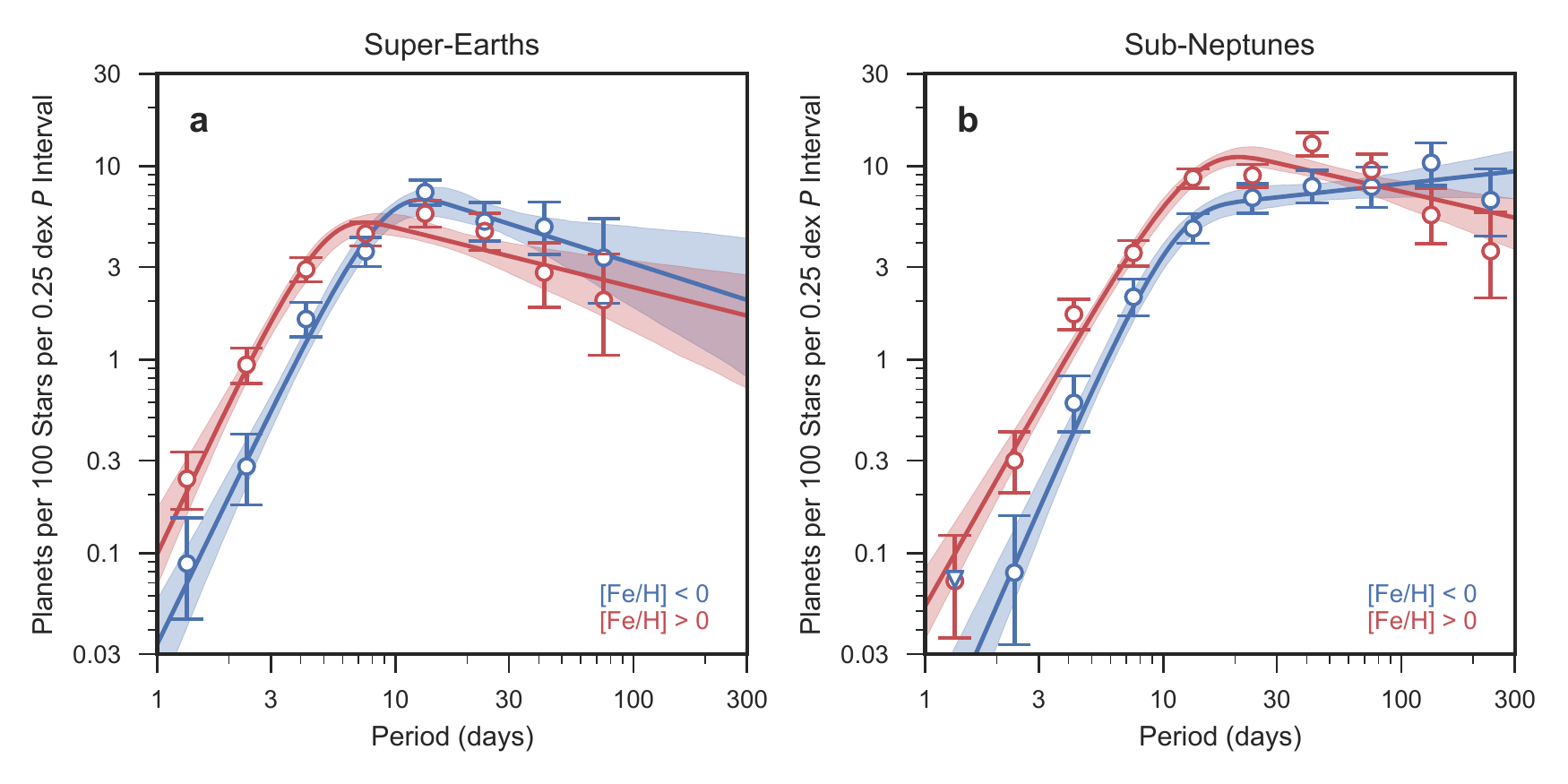}
\caption{Panel (a) is analogous to Figure~\ref{fig:occur-per}, except showing the period distribution of super-Earths at  different host star metallicities. For example, red points show the number of super-Earths per 100 stars with super-solar metallicity per 0.25~dex period interval. Hot super-Earths are more common around metal-rich stars. Panel (b) same as (a) except for sub-Neptunes. Hot sub-Neptunes are also more common around metal-rich stars.\label{fig:occur-per-smet}}
\end{figure*}

Figure~\ref{fig:occur-per-smet} shows the occurrence of super-Earths and sub-Neptunes as a function of orbital period for sub- and super-solar metallicity hosts. The occurrence rates for $P > 10$~days are generally consistent for the two metallicity bins. For $P < 10$~days, the occurrence rates of super-Earths/sub-Neptunes is $\approx$2--3 times higher for super-solar metallicity hosts compared to sub-solar metallicity hosts.

To quantify the metallicity correlation for our different planet classes, we modeled occurrence using the following parametric function:
\begin{equation}
\label{eqn:exp-per-smet}
\frac{d f}{d\log P\, d M } = C P^{\alpha} 10^{\beta M}.
\end{equation}
This model extends the following exponential model,
\begin{equation}
\label{eqn:exp-smet}
\frac{d f}{d M } = C 10^{\beta M},
\end{equation}
which was used in \cite{Fischer05} and in several subsequent works. Given that $M$ = $\log(n_{\mathrm{Fe}}/n_{\mathrm{H}}) - \log(n_{\mathrm{Fe}}/n_{\mathrm{H}})_{\odot}$, Equation~\ref{eqn:exp-smet} is equivalent to a power law relationship between planet occurrence and the number density of iron atoms in a star's photosphere relative to hydrogen,
\begin{equation}
\label{eqn:exponential}
\frac{d f}{d M } = C \left[\frac{n_{\mathrm{Fe}}/n_{\mathrm{H}}}{(n_{\mathrm{Fe}}/n_{\mathrm{H}})_{\odot}}  \right]^{\beta}.
\end{equation}

We fit the distribution defined in Equation~\ref{eqn:exp-per-smet} using the methodology from Section~\ref{sec:methodology-fitting} and list the associated model parameters in Table~\ref{tab:fit-smet}. 

For the hot super-Earths we found $\beta$ = \fit{persmet-hot-se-beta}, indicating a significant positive metallicity correlation. That $\beta \sim 1$ for hot super-Earths means that their occurrence is nearly proportional to the number of iron atoms in a star's photosphere, relative to hydrogen. Through our MCMC modeling, we found that the metallicity index $\beta$ was only weakly correlated with the period index $\alpha$. In contrast, $\alpha$ and the normalization constant $C$ were highly covariant.

The metallicity correlation steepens for larger planets. For hot sub-Neptunes, $\beta$ = \fit{persmet-hot-sn-beta}. For hot sub-Saturns and hot Jupiters, our planet sample \psamp contains only \samp{Hot Sub-Saturns n} and \samp{Hot Jupiters n} planets, respectively. Small sample size leads to larger uncertainties on $\beta$, and for hot sub-Saturns and Jupiters, $\beta$ = \fit{persmet-hot-ss-beta} and $\beta$ =  \fit{persmet-hot-jup-beta}, respectively. Despite these larger uncertainties, it is clear that hot sub-Saturns and Jupiters have significantly steeper metallicity dependencies than hot super-Earths and sub-Neptunes.

\begin{figure*}
\centering
\includegraphics{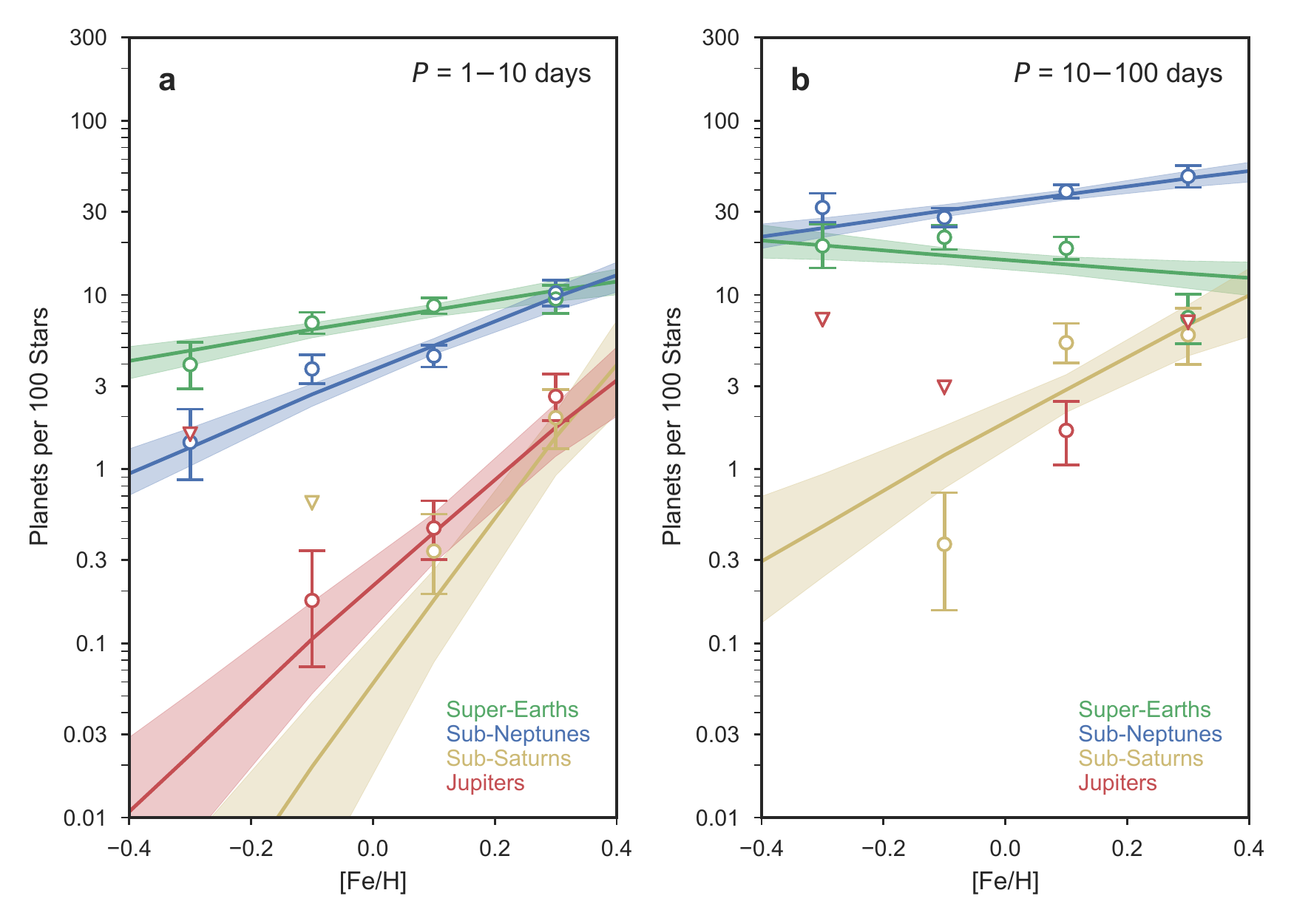}
\caption{Panel (a): Points show the number of planets per 100 stars for bins of host star metallicity spanning $\Delta M$ = 0.2~dex having $P$ = 1--10~days. Triangles represent upper limits (90\%). The colors correspond to different planet size classes. We modeled the observed occurrence rates with an exponential model, $df \propto P^\alpha 10^{\beta M} dM$, where $\beta$ characterizes the strength of a metallicity correlation (see Section~\ref{sec:occur-per-radius-smet}, Equation~(\ref{eqn:exp-per-smet})). The solid lines and bands show the best-fitting model and the $1\sigma$ credible range of models, respectively. Metallicity correlates with the occurrence of super-Earths ($\beta$ = \fit{persmet-hot-se-beta}), sub-Neptunes ($\beta$ = \fit{persmet-hot-sn-beta}), sub-Saturns ($\beta$ = \fit{persmet-hot-ss-beta}) and Jupiters ($\beta$ = \fit{persmet-hot-jup-beta}). The strength of the correlation increases with planet size.  Panel (b): same as (a) except for planets with $P$ = 10--100~days. Warm super-Earths are not correlated with metallicity ($\beta$ = \fit{persmet-warm-se-beta}). We observe positive correlations for larger planets. For warm sub-Neptunes ($\beta$ = \fit{persmet-warm-sn-beta}); for warm sub-Saturns ($\beta$ = \fit{persmet-warm-ss-beta}). Comparing the two panels, the metallicity correlation is stronger for $P$ < 10~days.\label{fig:occur-smet}}
\end{figure*}

Our occurrence rate model depends on both $P$ and $M$, and thus cannot be displayed on a 2D plot. For display purposes, we performed an integration from $P$~=~1--10~days.  Figure~\ref{fig:occur-smet} shows the number of planets belonging to different size classes per 100 stars having $P$ = 1--10~days for various metallicity intervals spanning $\Delta M$ = 0.2~dex. As in Figure~\ref{fig:occur-per}, the binned rates serve to guide the eye and are not used in the fitting.

We observe different metallicity dependencies for more distant planets. As shown in Figure~\ref{fig:occur-smet}, warm super-Earths are consistent with no metallicity dependence $\beta$~=~\fit{persmet-warm-se-beta}. Warm sub-Neptunes show a positive metallicity correlation, but their power law index $\beta$~=~\fit{persmet-warm-sn-beta} is significantly shallower than that of the hot sub-Neptunes. The warm sub-Saturns have a positive metallicity correlation of $\beta$ = \fit{persmet-warm-ss-beta}, again more shallow than that of the hot sub-Saturns. Finally, our sample contains too few warm Jupiters (\samp{Warm Jupiters n}) to search for a metallicity correlation.

In summary, some but not all planet subclasses are associated with stellar metallicity. Warm super-Earths exhibit no metallicity dependence. The planet-metallicity correlation steepens with increasing size and decreasing orbital period, with the hot Jupiters and hot sub-Saturns showing the steepest metallicity dependencies. We discuss some physical interpretation  for these trends in the following section.

\begin{deluxetable*}{llll}
\tablecaption{Best-fit Parameters for Planet-Metallicity Distributions\label{tab:fit-smet}}
\tablecolumns{4}
\tablehead{
\colhead{} & 
\colhead{$\log_{10} C$} & 
\colhead{$\alpha$} &
\colhead{$\beta$}
}
\tablewidth{0pt}
\startdata
Hot Super-Earth  & \fit{persmet-hot-se-logkp}  & \fit{persmet-hot-se-alpha}  & \fit{persmet-hot-se-beta}  \\
Warm Super-Earth & \fit{persmet-warm-se-logkp} & \fit{persmet-warm-se-alpha} & \fit{persmet-warm-se-beta} \\
Hot Sub-Neptune  & \fit{persmet-hot-sn-logkp}  & \fit{persmet-hot-sn-alpha}  & \fit{persmet-hot-sn-beta}  \\
Warm Sub-Neptune & \fit{persmet-warm-sn-logkp} & \fit{persmet-warm-sn-alpha} & \fit{persmet-warm-sn-beta} \\
Hot Sub-Saturn   & \fit{persmet-hot-ss-logkp}  & \fit{persmet-hot-ss-alpha}  & \fit{persmet-hot-ss-beta}  \\
Warm Sub-Saturn  & \fit{persmet-warm-ss-logkp} & \fit{persmet-warm-ss-alpha} & \fit{persmet-warm-ss-beta} \\
Hot Jupiter      & \fit{persmet-hot-jup-logkp} & \fit{persmet-hot-jup-alpha} & \fit{persmet-hot-jup-beta} \\
Warm Jupiter     & \nodata                     & \nodata                     & \nodata                    \\
\enddata
\tablecomments{Best-fit parameters associated the joint period-metallicity occurrence model defined in Equation~\ref{eqn:exp-per-smet}.}
\end{deluxetable*}

\section{Summary and Discussion}
\label{sec:discussion}
The \Kepler survey has revealed many valuable insights regarding the prevalence of planets with $P \lesssim$ 1~year. While numerous previous works have addressed \Kepler planet occurrence in the $P$--\Rp plane,%
\footnote{A non-exhaustive list includes: \cite{Youdin11,Catanzarite11,Traub12,Howard12,Fressin13,Dong13,Petigura13b,Foreman-Mackey14,Burke15,Dressing15,Mulders15}.}
we have leveraged the high-precision, high purity CKS catalog to produce a clearer picture of planet occurrence as a function of orbital period and size (Section~\ref{sec:occur-per-radius}; Figures~\ref{fig:checkerboard}--\ref{fig:contour}). This complements the work of \cite{Fulton17} that focused on super-Earth and sub-Neptune occurrence rates with $P$ = 1--100~days. Our occurrence rates may be easily incorporated into yield calculations for future surveys sensitive to planets with $P \lesssim 1$~yr. We found a hot Jupiter occurrence rate of $\samp{rate-hot-jup}$\%, which we compare to results from RV surveys in Section~\ref{sec:discussion-jovians}. While hot Jupiters are rare, we find that they occupy a distinct island in $P$--\Rp space, surrounded by a sea of still lower occurrence.

\begin{figure*}
\centering
\includegraphics{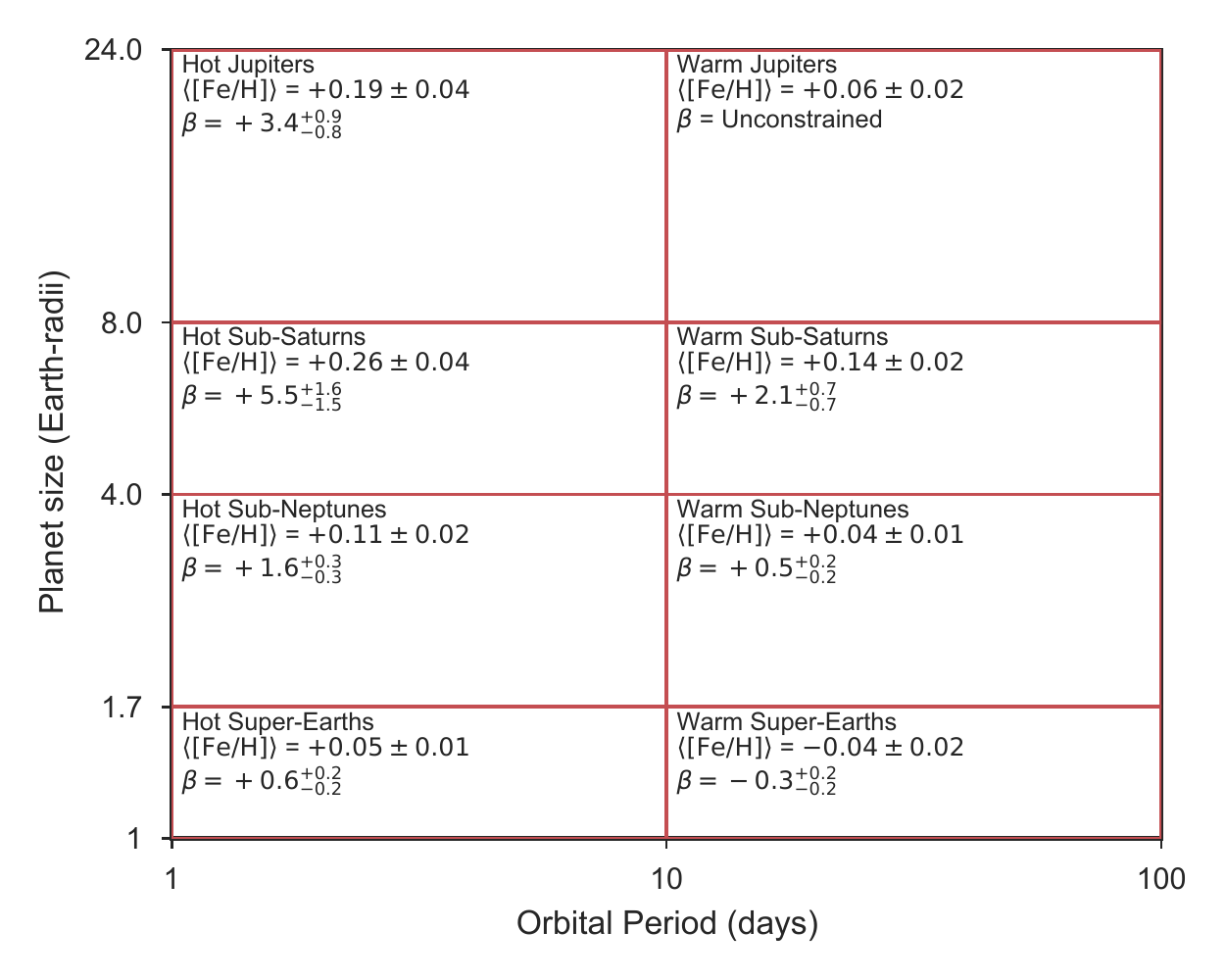}
\caption{Summary of the planet-metallicity correlation across the $P$--\Rp plane. For each planet subclass studied, we display the average host star metallicity \femn and the $\beta$ index, which quantifies the strength of the planet-metallicity correlation.\label{fig:summary}}
\end{figure*}

Using the CKS catalog combined with LAMOST metallicities for a large sample of \Kepler field stars, we have computed planet occurrence as a function of host star metallicity (Section~\ref{sec:occur-per-radius-smet}; Figures~\ref{fig:occur-per-smet} and \ref{fig:occur-smet}). We observe a clear association between stellar metallicity and the prevalence of certain types of planets. Figure~\ref{fig:summary} summarizes our findings: stellar metallicity is associated with the occurrence of hot planets ($P < 10$~days) and with the occurrence of large planets (\Rp > 1.7~\Re, $P < 100$~days). 

While the occurrence of some types of planets correlate strongly with metallicity, others exhibit only a moderate or negligible correlation. The interpretation of these metallicity trends is not yet clear.  There are many ways in which differences in metallicity might alter the processes of planet formation and evolution: through differences in the surface density of solids in the protoplanetary disk, the growth rate and abundance of solid cores, the availability of gas, the efficiency of gas accretion onto protoplanets, or the rate of disk migration. There may also be others. We offer some interpretations of these trends in Sections~\ref{sec:discussion-large}--\ref{sec:discussion-jovians}. Finally, we consider future studies that could extend our understanding of the connection between planets and host star metallicity in Section~\ref{sec:discussion-future}.

\subsection{Metallicity and the Prevalence of Large Planets}
\label{sec:discussion-large}
We found that metallicity is associated with the prevalence of large planets. This trend is consistent with the planet-metallicity correlation for Jovian-mass planets found by RV surveys \citep{Santos04,Fischer05} and with previous analyses of the metallicities of \Kepler planet hosts (e.g. \citealt{Buchhave12}). We observed that for smaller planets, the metallicity correlation weakens, which is consistent with trends observed in previous RV studies (e.g. \citealt{Sousa08,Ghezzi10}) and metallicity studies of \Kepler planet hosts (e.g. \citealt{Buchhave12,Buchhave14}). Our analysis builds upon these previous studies by providing planet occurrence as a function of metallicity based on the high-precision, high purity CKS catalog of planet radii and metallicities.

There are roughly 15 warm super-Earths per 100 stars over the metallicities ranging from $-$0.4~dex to $+$0.4~dex. If we assume that stellar metallicity traces disk solid surface densities from 0.10--0.40~au (which correspond $P\approx$~10--100~day orbits), then warm super-Earths can form with moderate efficiency, even in disks with low solid surface densities. Moreover, a factor of 6 increase in solids does not enhance the likelihood of a star to host a warm super-Earth. 

RV and TTV mass measurements of super-Earths typically find masses of 1--10~\Me (\citealt{Marcy14}, see Figure 7 of \citealt{Sinukoff17b} for an updated census). The bulk densities of these planets are usually consistent with combinations of
rock and iron, but are inconsistent with even small envelope fractions of $M_{\mathrm{env}} / \Mp \sim 1$\%, which would significantly reduce the observed bulk densities.

Even though sub-Neptunes are significantly larger than super-Earths, they share several important characteristics. Sub-Neptunes are typically 5--15~\Me, which overlaps with the super-Earth mass range. Sub-Neptune bulk densities require H/He envelopes that are 1--10\% of the planet's mass, which significantly enlarge the planets without significantly altering their masses. Therefore, the typical sub-Neptune contains a comparable amount of solid material as a typical super-Earth. Also, their overall occurrence rates are roughly comparable at 5--10\% per 0.25~dex period interval (Figure~\ref{fig:occur-per-smet}).

We see an important difference between super-Earths and sub-Neptunes when we compare their dependencies on stellar metallicity. There are comparable numbers of warm super-Earths and warm sub-Neptunes around stars having \fe = [$-$0.4,$-$0.2]~dex, about 20 per 100 stars. Unlike the warm super-Earths, the occurrence of warm sub-Neptunes increases with metallicity. There are about 40 warm sub-Neptunes per 100 stars having \fe = [$+$0.2,$+$0.4]~dex. 

Perhaps the enriched disks of metal-rich stars form super-Earths more efficiently, but they easily acquire the few percent envelopes and are converted into sub-Neptunes. This could occur if super-Earths form more quickly in metal-rich disks, allowing for more time to accrete 1--10\% gaseous envelopes.

Such a timing argument was proposed by \cite{Dawson15}, to explain an apparent absence of warm super-Earths around the highest metallicity stars (\fe > 0.18~dex) in the \cite{Buchhave14} metallicity catalog.%
\footnote{The \cite{Dawson15} definition of warm super-Earths was slightly different than ours: $\Rp < 1.5$~\Re, $P < 15$~days.}
While only 12/132 (9\%) of our warm super-Earths have \fe > 0.18~dex, this is consistent with the fact that only 12\% of field stars have \fe > 0.18~dex. As shown in Figures \ref{fig:per-prad-slices-equal-stars} and \ref{fig:occur-smet}, we do not observe such an absence of warm super-Earths, but instead find nearly equal occurrence of warm super-Earths around stars of wide-ranging metallicities. This underscores the importance of having accurate knowledge of the distribution of field star metallicities in planet-metallicity studies. Nevertheless, accelerated formation of cores in metal-rich disks may explain the rise in warm sub-Neptune occurrence with metallicity.

An alternative to the timing argument, discussed above, is that high disk metallicities may  somehow increase the rate of envelope accretion. However, the gas accretion models of  \cite{Pollack96} and (more recently) \cite{Lee15a} predict the opposite effect: dusty envelopes should accrete more slowly, due to higher opacities and longer cooling timescales.

As we consider larger warm planets, the metallicity correlation becomes stronger. Sub-Saturns have a $\beta$ index of \fit{persmet-warm-ss-beta}. If disk metallicity assists with the accretion of gas through accelerated core formation, it may also explain the sub-Saturns. However, abrupt strengthening of the metallicity correlation above 4~\Re and the roughly order of magnitude decrease in the frequency of sub-Saturns compared to sub-Neptunes suggests the formation pathways of sub-Saturns and sub-Neptunes may be quite different. 

Around 20 sub-Saturns have well-measured masses from either RVs or TTVs (see \citealt{Petigura17a} for a recent compilation). For sub-Saturns, we observe an order of magnitude scatter in the observed masses at a given size, indicating a diversity in core and envelope masses. While some sub-Saturns have $\approx$5~\Me cores, similar to the super-Earths and sub-Neptunes, many have cores of $\approx$50~\Me. In addition, the most massive sub-Saturns tend to be found around the most metal-rich hosts \citep{Petigura17a}.

The existence of $\approx$50~\Me cores in planets with $20\%$ envelope fractions challenges the classic core-accretion models of \cite{Pollack96} that predict that cores larger than 10~\Me should undergo runaway accretion. Perhaps these massive sub-Saturn cores are the result of the late-stage mergers (or series of mergers) of 10~\Me cores. This formation scenario requires one or more closely spaced 10~\Me planets. As we have shown, the probability for a star to produce a $10$~\Me core increases with metallicity. Therefore, the probability for a star to produce two $10$~\Me cores likely increases with a steeper power law index. Thus, the production of sub-Saturns by collisions may explain the mass-metallicity dependence and the steeper relationship between metallicity and planet occurrence. This could also explain the mass-metallicity correlation. This theory also predicts that if late-stage mergers play a large role in the formation of sub-Saturns, they should produce relic eccentricities that are observable at later times.

Our planet sample includes only four warm Jupiters, which is insufficient to search for trends. \cite{Fischer05} analyzed 1040 FGK-type stars from the Keck, Lick, and Anglo-Australian Telescope planet search programs and found a planet-metallicity correlation with a $\beta = 2$ index. However, this sample included planets with $P$ = 1--4000~days, with the bulk of the sample having $P > 300$~days longer than the periods considered here.

\subsection{Metallicity and the Prevalence of Short-period Planets}
\label{sec:discussion-close}

We found that metallicity is associated with the presence of short-period planets having $P$ < 10~days. This is consistent with trends observed by \cite{Mulders16} and \cite{Dong17}, who studied the host star metallicities for samples of 665 and 295 planets, respectively. Both studies used spectroscopically constrained metallicities from LAMOST. Our analysis incorporates the high-precision CKS metallicities for a larger sample of $\samp{n-cand-cks-pass}$ planets.

While close-in planets, of all sizes, are rare compared to more distant planets, their frequency increases as a function of stellar metallicity. In Section~\ref{sec:occur-per-radius}, when considering just the period distribution of super-Earths and sub-Neptunes, we found that for $P < 10$~days, the occurrence of super-Earths and sub-Neptunes declines with decreasing period. The slope of this falloff is approximated by $d f \propto P^{\alpha} d \log P$ with $\alpha$ values of \fit{per-all-se-gamma+beta} and \fit{per-all-sn-gamma+beta}, respectively.

In Section~\ref{sec:occur-per-radius-smet}, when considering the joint $P$--\fe occurrence distribution, we found that both hot super-Earths and sub-Neptunes are correlated with host star metallicity. This is especially noteworthy for the hot super-Earths, given that the warm super-Earths exhibit no such correlation. For $P < 10$~days, the frequency of hot super-Earths grows from 4\% for stars having \fe = [$-0.4$,$-0.2$]~dex to 10\% for stars having \fe = [$+0.2$,$+0.4$]~dex (see Figure~\ref{fig:occur-smet}). We see an even steeper increase in the hot sub-Neptune rates from 1\% to 8\% for the same metallicity intervals. In Sections~\ref{sec:discussion-insitu} and \ref{sec:discussion-migration}, we consider two possible formation pathways: (1) {\em in situ} models, where planets form near their final location, and (2) high eccentricity migration, where the semi-major axes of fully formed planets are altered by major scattering events. In Section~\ref{sec:discussion-tests}, we consider some observational tests that may distinguish between these scenarios.

\subsubsection{In Situ Formation}
\label{sec:discussion-insitu}
The prevalence of compact multi-planet systems of super-Earths/sub-Neptunes discovered by \Kepler helped inspire {\em in situ} planet formation models (see, e.g., \citealt{Hansen13} and \citealt{Chiang13}). \cite{Chiang13} introduced the minimum mass extrasolar nebula (MMESN) as a means of estimating the surface density profile $\Sigma(a)$ of the protoplanetary disk. They took a sample of \Kepler planets, estimated the mass in solids, and smeared those solids out over an area $2\pi a^2$. The MMESN assumes no large-scale migration of solid building blocks or planets. In the context of the MMESN framework, the falloff in super-Earth and sub-Neptune occurrence at $P < 10$~days reflects a declining surface density profile. \cite{Lee17} offered a physical explanation for declining density profiles. In their models, disks are truncated by interactions with the protostar magnetosphere at the co-rotation radius.

The existence of hot sub-Neptunes presents a challenge to strictly {\em in situ} models, because sub-Neptunes with $P$ < 10~days are vulnerable to photo-evaporation. \cite{Owen17} performed a population synthesis starting with 3--10~\Me cores and a period distribution of $d f  \propto P^{1.9} d \log P$, which is similar to the power law index presented in this work. They then gave the simulated planets a range of envelope fractions from 1 to 30\% and tracked the radii of these planets as they were subjected to photo-evaporation. The \cite{Owen17} simulations produced very few sub-Neptunes with $P < 10$~days.

In the \cite{Lee17} model, hot super-Earth/sub-Neptune occurrence rates are set by stellar rotation on the pre-main sequence. It is not clear if stellar metallicity is correlated with pre-main-sequence rotation rates, and \cite{Lee17} did not predict a metallicity dependence for hot super-Earths/sub-Neptunes. 

Metallicity may still influence the vertical and radial profiles of the inner disk, even if there is no correlation with stellar rotation. The inner regions of high metallicity disks may have higher densities of free electrons, which could result in stronger coupling to stellar magnetic fields and change the accretion rate and density profile of the disk. The formation of hot super-Earths/sub-Neptunes may also involve significant radial transport of solids, either through the inward drift of ``pebbles'' that are no longer aerodynamically coupled to the gas disk (see, e.g., \citealt{Lambrechts17}, and references therein). The efficiency of this process is also governed by the radial and vertical density profiles.

Future theoretical modeling of protoplanetary disks is needed to investigate whether {\em in situ} models can reproduce the metallicity dependence observed here. {\em In situ} models tend to produce dynamically cool planetary systems, which can be corroborated by present-day eccentricity and obliquity measurements 

\subsubsection{High Eccentricity Migration}
\label{sec:discussion-migration}
High eccentricity migration may explain the hot super-Earth/sub-Neptune period and metallicity distributions. Some of the hot super-Earths/sub-Neptunes may be the result of scattering events between multiple planets. Such encounters preserve the total orbital energy of both planets, but the semi-major axes of both planets may be significantly altered. \cite{Dawson13} proposed a similar mechanism to explain the fact that high eccentricity giant planets with $a$ = 0.1--1.0~au, almost exclusively orbit stars with super-solar metallicities.

A formation pathway involving scattering may help to explain the existence of hot sub-Neptunes, which are at risk of photo-evaporative stripping (see Section~\ref{sec:discussion-insitu}). Scattering events may occur late enough such that a sub-Neptune is not subjected to the majority its star's XUV output ($t \gtrsim 100$~Myr). 

For hot sub-Neptunes, the metallicity correlation is stronger ($\beta$ = \fit{persmet-hot-sn-beta}) than for the warm sub-Neptunes ($\beta$ = \fit{persmet-warm-sn-beta}). This steeper relationship may support a scattering interpretation given that the probability of scattering depends on the probability of a disk producing two closely spaced planets of comparable mass. If high eccentricity migration is active, the present-day systems planetary systems should be dynamically hot, provided that star--planet interactions are not strong enough to circularize or realign orbits. Again, this may be corroborated with eccentricity and obliquity measurements.

\subsubsection{Observational Tests}
\label{sec:discussion-tests}
Additional clues to the formation of hot super-Earths/sub-Neptunes may lie in their orbital eccentricities and degree of alignment with stellar spin axes (obliquities). If planets initially form in the plane of their disks, but are then scattered by other planets, we predict relic eccentricities and obliquities. If these planets form in relative isolation, and are simply a product of the local disk density profile, their eccentricities and obliquities should be low.

Current state-of-the-art RV facilities have provided mass measurements of several dozen hot super-Earths and sub-Neptunes, but are not precise enough to measure orbital eccentricities well. Obliquity measurements via the Rossiter--McLaughlin technique have been made for only a handful of sub-Jovian-size planets (e.g. HAT-P-11, \citealt{Winn10c,Hirano11}). As the precision and observing efficiency of RV facilities improve, we should be able to place tighter constraints on the orbits of these hot planets.

\subsection{Hot Jupiters}
\label{sec:discussion-jovians}
Hot Jupiters represent an extreme of planet formation, and many theories have been proposed to explain their formation. A non-exhaustive list of theories includes star--planet Kozai (e.g. \citealt{Wu03}), planet-planet Kozai (e.g. \citealt{Naoz11}), smooth disk-driven Type-I migration (e.g. \citealt{Ida08}), high eccentricity migration, and {\em in situ} formation (e.g. \citealt{Batygin16}). Hot Jupiters are also of interest because they offer a point of comparison between the \Kepler population of planets and those around nearby stars. 

We find a hot Jupiter rate of $\samp{rate-hot-jup}\%$,%
\footnote{The rate of $\samp{rate-hot-jup}\%$ is slightly higher than a simple sum of the most likely rates from each hot Jupiter bin from Figure~\ref{fig:checkerboard} due to the asymmetric uncertainties from low counts per bin.}
which is about $1.5\sigma$ larger than $0.4\pm0.1\%$ reported by \cite{Howard12}. This difference cannot be due to differences the treatment of completeness; for hot Jupiters, \pdet is nearly unity.

The larger occurrence rate presented here is likely due to the improved radius measurements in the CKS sample. \cite{Howard12} included 13 hot planets with \Rp = 5.6--8.0~\Re, slightly too small to be classified as a hot Jupiter by their criteria. With our improved CKS planet radii, we have found that these large, hot sub-Saturns are exceptionally rare. Our planet sample \psamp includes just 3 such planets, while \cite{Howard12} included 13, even though their stellar parent is less than twice as large as ours \nssamp (58041 vs. \samp{n-stars-field-pass-eff}). Some of the hot sub-Saturns in \cite{Howard12} were misidentified hot Jupiters.

A major challenge in comparing the hot Jupiter rates in the \Kepler and the solar neighborhood is their low intrinsic occurrence. Even though our planet sample \psamp was drawn from a parent stellar sample \ssamp containing \samp{n-stars-field-pass-eff} stars, \psamp contained just \samp{Hot Jupiters n} detections due to low intrinsic occurrence and the requirement of transiting geometries. Even blind RV surveys which are not limited to transiting geometries must observe hundreds of stars to detect a few hot Jupiters. Another challenge is that RV surveys are typically not magnitude-limited samples like \ssamp. RV surveys often prioritize stars that are single, slowly rotating, and have stellar properties that fall within a preselected range of interest (e.g. M-stars or evolved stars). \cite{Wright12} attempted to retroactively remove these selection effect from the California Planet Search database, to produce a magnitude-limited sample of stars that could be directly compared to the \Kepler stellar sample. \cite{Wright12} reported 10 hot Jupiters from a sample of 836 stars or a rate of $1.2\pm0.4\%$, twice the rate presented here. However, the large fractional uncertainty in the \cite{Wright12} measurement means that these two results differ by only 1.5$\sigma$.

In Section~\ref{sec:occur-per-radius-smet}, we modeled the hot Jupiter occurrence as $df \propto P^{\alpha} 10^{\beta M} d \log P d M$ and found a $\beta$ index of \fit{persmet-hot-jup-beta}. This strong association between hot Jupiters and metallicity has also been observed by RV stars. \cite{Fischer05} studied the occurrence of Jupiter-mass planets in the CPS sample and modeled their occurrence according to $df \propto 10^{\beta M}$ and $\beta$ index of 2.0. \cite{Johnson10} performed a similar analysis, considered an additional mass dependence $df \propto  \Mstar^{\alpha} 10^{\beta M}$ and found a $\beta$ index of $1.2\pm0.2$. It is worth recalling that both studies included long period planets ($P > 1$~yr), which constituted the bulk of the sample. \cite{Guo17} repeated the \cite{Johnson10} analysis, but restricted the study to hot Jupiters and found $\beta$ = $2.1\pm0.7$. While this result differs with our measurement of $\beta$ by about 2$\sigma$, it is clear that hot Jupiters in the solar neighborhood and in the \Kepler field are both strongly associated with metallicity. 

In the previous sections, we have noted the general tendency for the metallically correlation to steepen with decreasing orbital period and increasing planet size, and have hypothesized some processes that could account for these trends such as accelerated core/envelope growth, higher disk densities, and planet-planet scattering. Given this general trend, it is perhaps not surprising that hot Jupiters would have the strongest metallicity correlation. However, given that the hot Jupiters occupy a distinct island on the $P$--\Rp plane, it is possible that their formation pathway is radically different than that of the other planet classes considered here.

\subsection{Future Survies}
\label{sec:discussion-future}
Our target sample contains very few stars (\samp{n-stars-cks-pass-smet<-0.4}) with metallicities below $-0.4$~dex. As a result, we are insensitive to planet occurrence at low metallicities. Probing planet occurrence at low metallicities would shed light on some of the questions raised by this analysis, such as, What is the minimum metallicity needed to form a warm super-Earth?

Since metal-poor stars are rare, a blind survey like \Kepler must target a large number of stars in order to capture sufficient numbers of low metallicity stars. There is some room for improvement with the existing \Kepler sample. Our sample was drawn from $\approx$38,000 of the $\approx$150,000 stars observed by \Kepler due to our magnitude limit. Characterizing the metallicities of \Kepler planet hosts and field stars fainter than \kepmag = 14.2~mag would enlarge the sample of metal-poor stars. However, given that these new stars would be between \kepmag = 14--16~mag, their amenability to the detection of small planets is poor.

Upcoming missions like {\em TESS} \citep{Ricker15} and {\em PLATO} \citep{Rauer14} will survey more stars than \Kepler and thus cast a significantly larger net for rare metal-poor stars. {\em Gaia} will also identify metal-poor stars across the sky. A targeted RV survey of low metallicity stars would also probe planet occurrence at low metallicities. Such a survey would not be restricted to transiting planets, and would need to survey fewer stars to gather statistically significant planet samples. 

Finally, one could search for planets in globular clusters. \cite{Gilliland00} conducted an important early transit search with HST, which targeted 47 Tucanae (\fe = $-$0.78~dex). The expected yield was 17 hot Jupiters, but the survey yielded no detections, which was attributed to low metallicities. Recently, \cite{Masuda17} re-assessed the significance of the null result, using updated knowledge of the typical size, periods, and host star properties of hot Jupiters. Their revised yield calculation was $2\pm1$, and thus the null result does not require a metallicity effect. However, based on the steep dependence of hot Jupiter occurrence with metallicity, yields would likely be low. Future globular cluster surveys with higher precision or longer baselines would help constrain planet population at $-1$~dex. However, high stellar densities of globulars may complicate direct comparisons to field stars.

\section{Conclusion}
\label{sec:conclusion}
We have measured planet occurrence as a function of period, radius, and stellar metallicity within a magnitude-limited sample of \Kepler target stars. We have leveraged precise planet radii and stellar metallicities from the CKS catalog of \Kepler planet-hosting stars, as well as the metallicity distribution of \Kepler field stars measured by LAMOST. In summary, for $P < 100$~days the default planetary system contains either no planets detectable by \Kepler (e.g. the solar system) or a system of one or more super-Earths/sub-Neptunes with $P$ = 10--100~days. Stars with high metallicity are associated with some mechanism(s) that also allows for ``misplaced'' planets: sub-Saturns/Jupiters with $P$ < 100~days and super-Earths/sub-Neptunes with $P$ < 10~days. We have noted the general features of planet occurrence as a function of $P$, \Rp, and \fe and have offered some speculation regarding their physical origins. Detailed population synthesis models that treat the disk profile, growth of cores, migration, dynamical instabilities, and photo-evaporation must produce these features.

\acknowledgments
The CKS project was conceived, planned, and initiated by A.W.H., G.W.M., J.A.J., H.T.I., and T.D.M.  
A.W.H., G.W.M., J.A.J. acquired Keck telescope time to conduct the magnitude-limited survey.  
We thank the many observers who contributed to the measurements reported here.


We thank the referee, Lars Buchhave, for his detailed and thoughtful comments on the techniques and interpretations. We also thank Konstantin Batygin, Brendan Bowler, Ian Crossfield, and Eve Lee for enlightening conversations that improved the final manuscript. 

\Kepler was competitively selected as the tenth NASA Discovery mission. Funding for this mission is provided by the NASA Science Mission Directorate. We thank the Kepler Science Office, the Science Operations Center, the Threshold Crossing Event Review Team (TCERT), and the Follow-up Observations Program (FOP) Working Group for their work on all steps in the planet discovery process, ranging from selecting target stars and pointing the Kepler telescope to developing and running the photometric pipeline to curating and refining the catalogs of Kepler planets.

The Guoshoujing Telescope (the Large Sky Area Multi-Object Fiber Spectroscopic Telescope, LAMOST) is a National Major Scientific Project built by the Chinese Academy of Sciences. Funding for the project has been provided by the National Development and Reform Commission. LAMOST is operated and managed by the National Astronomical Observatories, Chinese Academy of Sciences.  

E.A.P. acknowledges support from Hubble Fellowship grant HST-HF2-51365.001-A awarded by the Space Telescope Science Institute, which is operated by the Association of Universities for Research in Astronomy, Inc., for NASA under contract NAS 5-26555.
L.M.W. acknowledges support from Gloria and Ken Levy and from the Trottier Family.
T.D.M. acknowledges NASA grant NNX14AE11G.

This work made use of NASA's Astrophysics Data System Bibliographic Services.

Finally, the authors wish to recognize and acknowledge the very significant cultural role and reverence that the summit of Maunakea has always had within the indigenous Hawaiian community.  We are most fortunate to have the opportunity to conduct observations from this mountain.

\software{All code used in this paper is available at \url{https://github.com/California-Planet-Search/cksmet/}. We made use of the following publicly available Python modules: astropy \citep{Astropy-Collaboration13}, isoclassify \citep{Huber17}, lmfit \citep{Newville14}, matplotlib \citep{Hunter07}, numpy/scipy \citep{numpy/scipy}, pandas \citep{pandas}, and pinky (\url{https://github.com/pgromano/pinky}).}


\bibliography{manuscript.bib}

\appendix
\section{Calibration of LAMOST Metallicities}
\label{sec:calibrate-lamost}
This paper relies on CKS metallicities for the planet-hosting stars and LAMOST metallicities to characterize the metallicity distribution of \Kepler field stars. Differences in metallicity scales often result from different spectral datasets or spectroscopic pipelines. The CKS and LAMOST surveys used different spectra, line-lists, and model-atmospheres. Here, we assess the agreement between the CKS and LAMOST metallicity scales, correct for a small offset, and report the uncertainty in our calibration.

\begin{figure*}
\centering
\includegraphics[width=1.0\textwidth]{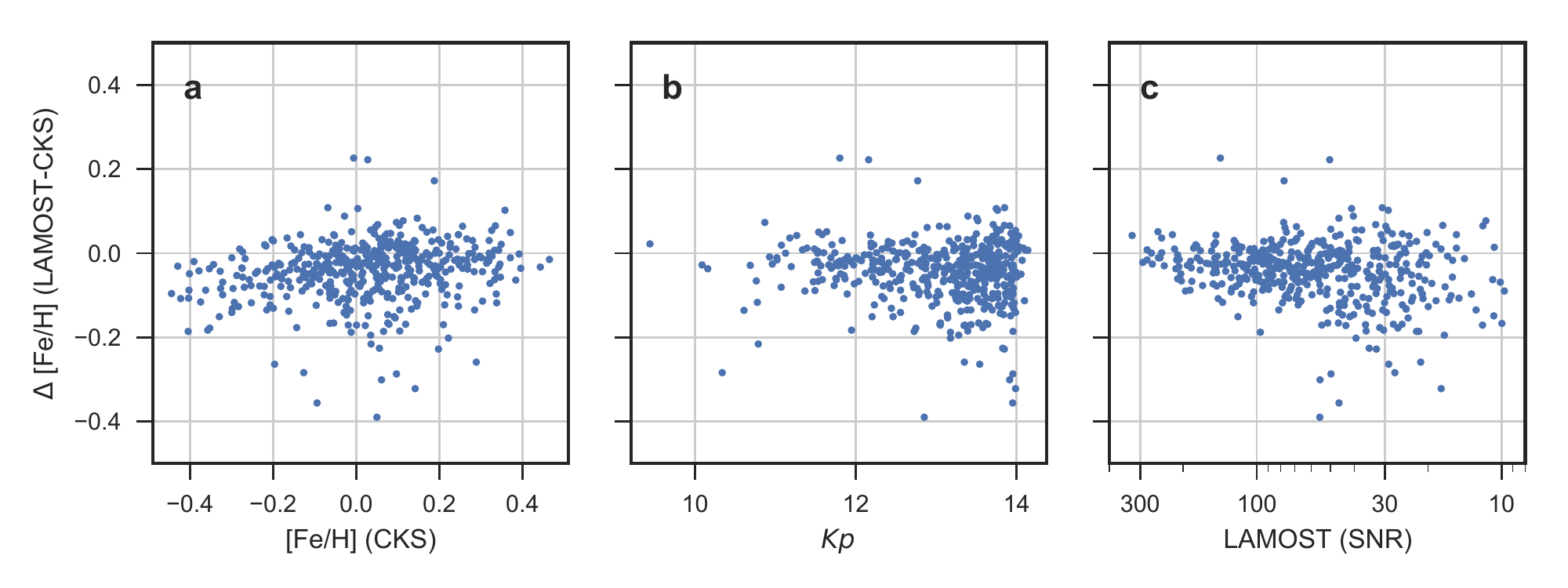}
\caption{Difference between CKS and LAMOST \fe for \samp{n-cks-lamo-tot} stars in common as a function of CKS metallicity (a), \kepmag (b), and the reported S/N of the LAMOST spectra (c). We observe a small metallicity-dependent systematic difference, which we calibrate out in Appendix~\ref{sec:calibrate-lamost}.\label{fig:lamo-diff}}
\end{figure*}

There are \samp{n-cks-lamo-tot} stars observed by both CKS and LAMOST. Figure~\ref{fig:lamo-diff} shows the difference between LAMOST and CKS metallicities ($\Delta \fe$) as a function of CKS metallicity, \kepmag, and the S/N of the LAMOST spectra. On average, the LAMOST metallicities are 0.04~dex lower and we observe a small metallicity-dependent systematic trend. The dispersion of $\Delta \fe$ increases for fainter stars or decreasing LAMOST S/N (the CKS spectra have homogeneous S/N). This indicates that the precision of the LAMOST metallicities declines with decreasing S/N.

We derived a correction $\Delta$ that calibrates the LAMOST metallicities onto the CKS scale via $\mathrm{LAMOST}_\mathrm{cal} = \mathrm{LAMOST}_\mathrm{raw} + \Delta$. The correction is a linear function of the following form:
\begin{eqnarray}
\Delta \fe  & = & c_0 + c_1 \left(\frac{\fe}{0.1~\dex}\right).
\end{eqnarray}
The coefficients are chosen such that they minimize the RMS difference between the calibrated LAMOST and CKS parameters. Before fitting, we removed \samp{n-cks-lamo-out} stars where the CKS and LAMOST metallicities differed by more than 0.2~dex. These outliers are likely due to rare failure modes of the LAMOST pipeline.%
\footnote{To verify that our calibration does not depend sensitively on our choice to remove  outliers, we left them in but solved for the coefficients that minimized the sum of the absolute differences, a metric which is resistant to outliers. The coefficients agreed to 1.5$\sigma$ or better.}

\begin{figure*}
\centering
\includegraphics[width=0.8\textwidth]{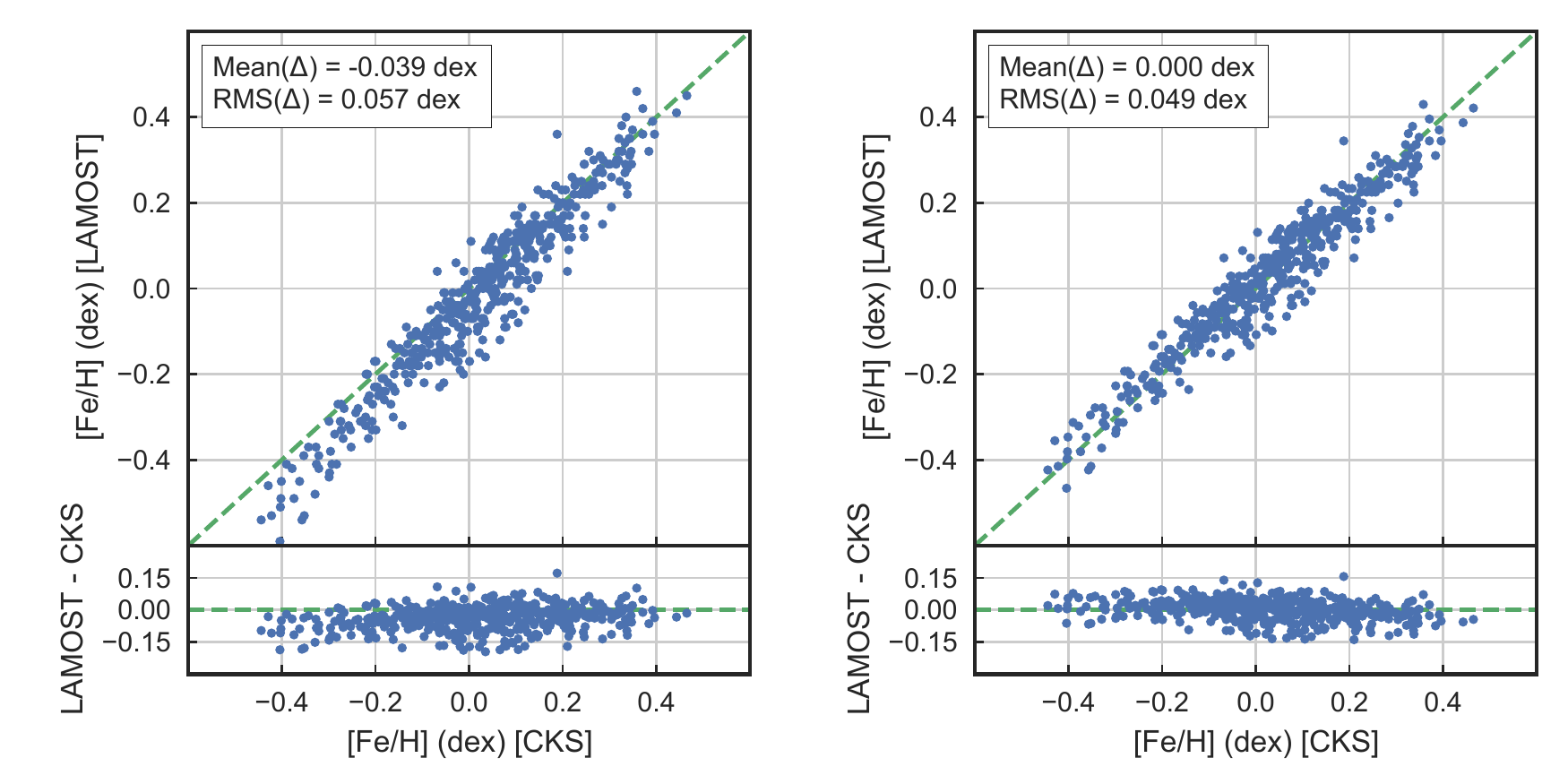}
\caption{Comparison of CKS and LAMOST metallicities. (Left) Uncalibrated LAMOST metallicities as a function of CKS metallicities for stars in common. The green dashed line represents equality. On average, the LAMOST metallicities are 0.05~dex lower than the CKS values. (Right) comparison of CKS and LAMOST metallicities after removing a linear systematic trend.\label{fig:lamo-on-cks}}
\end{figure*}

We estimated the uncertainties on the best-fit parameters using bootstrap resampling (with replacement), as described in \cite{Press02}. The best-fit coefficients are $c_0 = \samp{L2-c0} \pm \samp{L2-c0-err}$ and $c_1 = \samp{L2-c1} \pm \samp{L2-c1-err}$. We show a one-to-one comparison of the CKS and LAMOST metallicities (without the \samp{n-cks-lamo-out} outliers) in Figure~\ref{fig:lamo-on-cks}. After placing the LAMOST metallicities on the CKS scale, there is no residual mean offset (by construction) and a RMS dispersion of 0.05~dex.

Our calibration amounts to a correction of $\samp{L2-val-m0.5}$~dex at \fe = $-$0.5~dex and $\samp{L2-val-p0.5}$~dex at \fe = $+$0.5~dex. The uncertainty associated with our correction at $-$0.5~dex and $+$0.5~dex sets an upper bound on the residual systematic offset between the LAMOST and CKS metallicity scales relevant to this paper, which treats planet hosts having metallicities between $-$0.5 and $+$0.5~dex. We estimate that the CKS and the calibrated LAMOST metallicities are consistent to 0.01~dex.

To verify that the main results of this paper are not sensitive to zero-point offsets between the CKS and LAMOST metallicities scales, we performed a parallel analysis after perturbing our calibrated LAMOST metallicities by 0.01 dex. We compared the slopes of the metallicity distributions for different planet classes, which is quantified by the $\beta$ index in Equation~\ref{eqn:exp-per-smet}. Adding 0.01 dex to the LAMOST metallicities increases \ntrial in the high metallicity bins, which results in lower measured occurrence at high metallicities. For low metallicity bins, \ntrial decreases, resulting larger measured occurrence. The net effect is that the slope of the metallicity distribution becomes less steep (i.e., smaller $\beta$ indices). (Subtracting 0.01 dex from the calibrated LAMOST metallicities has the opposite effect and results in larger beta-indices.)
 
\begin{figure*}
\centering
\includegraphics{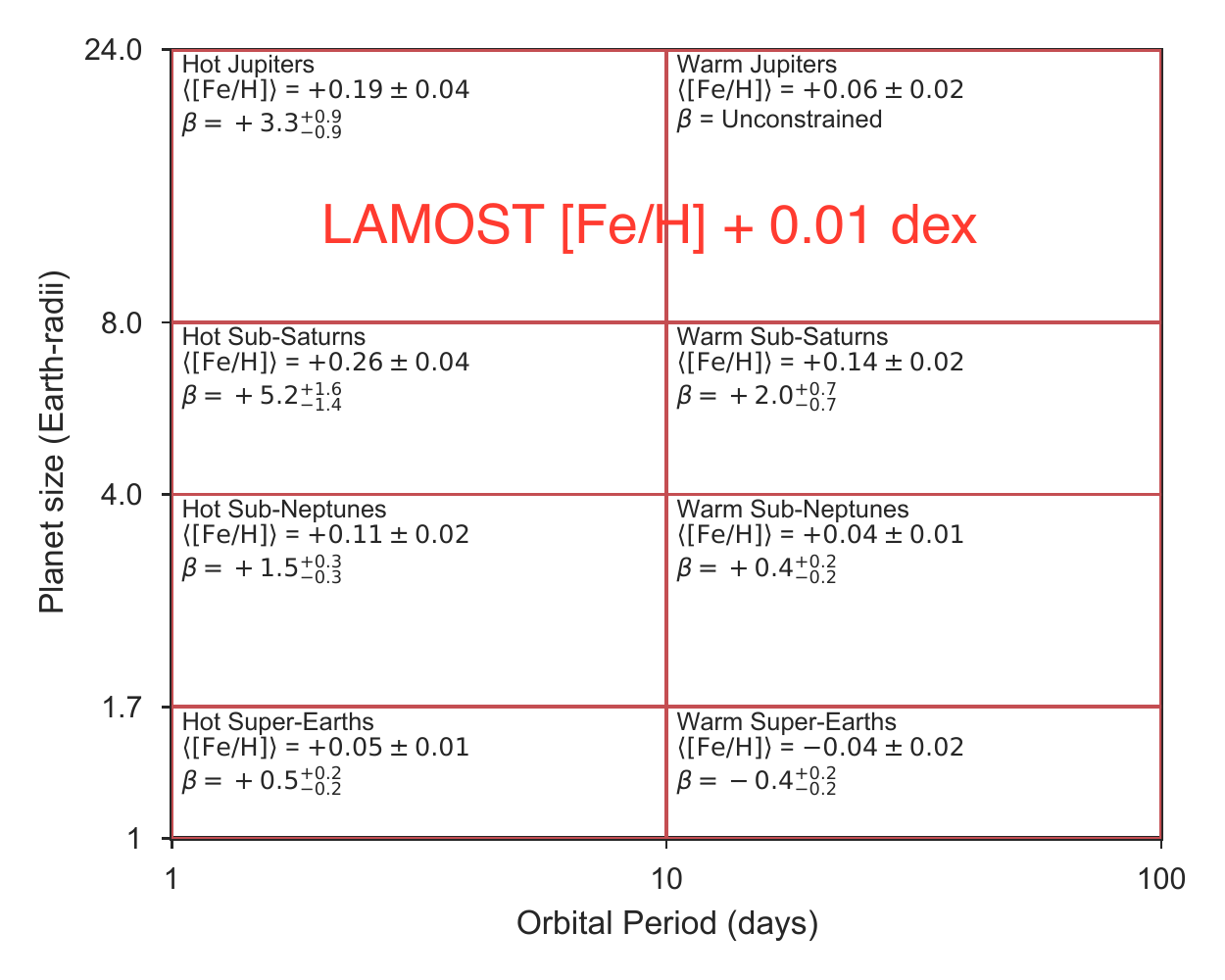}
\caption{Same as Figure~\ref{fig:summary}, but after adding 0.01~dex to the LAMOST metallicities to simulate the effect of residual offsets between the CKS and LAMOST metallicity scales.  For all planet classes, the change in $\beta$ is smaller than the adopted uncertainty on $\beta$, which incorporates counting statistics alone. The uncertainties on $\beta$ are therefore dominated by counting statistics, rather than uncertainties in our CKS-LAMOST metallicity calibration. \label{fig:summary+001}}
\end{figure*}

Figure~\ref{fig:summary+001} is analogous to Figure~\ref{fig:summary}, but we created it after shifting the LAMOST metallicities by +0.01 dex. Comparing the $\beta$ indices, these two figures summarize the effects of a possible offset in metallicity scales. For all planet classes, $\beta$ decreases, as expected. For example, for warm super-Earths, $\beta$ decreases from $-0.3\pm0.2$ to $-0.4\pm0.2$; for warm sub-Neptunes, $\beta$ decreases from $+0.5\pm0.2$ to $+0.4\pm0.2$. For all planet classes, the change in $\beta$ is smaller than our adopted $1\sigma$ uncertainties from our maximum-likelihood fitting, which only incorporates counting statistics. Therefore, we conclude that the uncertainties on the $\beta$ indices due to our LAMOST-CKS calibration are smaller than those due to counting statistics. Thus, the conclusions of this paper are not sensitive to errors in our LAMOST-CKS metallicity calibration.

\section{Metallicity and Planet Detectability}
\label{sec:smet-detectability}
In Section~\ref{sec:smet}, we compared the properties of planets belonging to host stars of different metallicities, and in Section~\ref{sec:occur-per-radius-smet}, we computed planet occurrence as a function of host star metallicity. For both calculations, we made the assumption that planet detectability is not a strong function of host star metallicity. Here, we check the validity of such an assumption. 

\begin{figure*}
\centering
\includegraphics{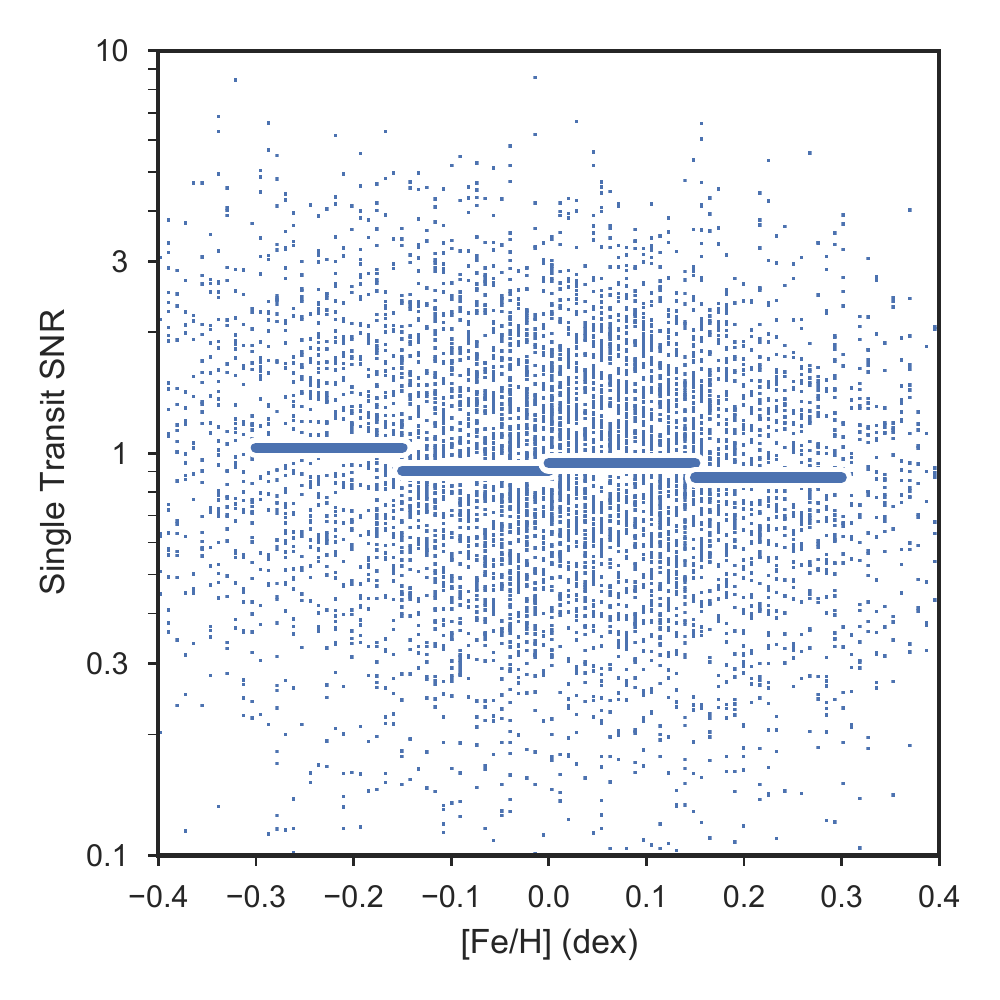}
\caption{Planet detectability vs. metallicity. The y-axis shows the signal-to-noise ratio of a putative 1~\Re planet, which has a transit duration of three hours. We do not observe a significant trend in planet detectability with metallicity.\label{fig:smet-detect}}
\end{figure*}

Metallicity could correlate with planet detectability through a correlation with stellar size or noise properties. As a direct measure of the possible dependence of planet detectability with metallicity, we computed the single transit signal-to-noise ratio (S/N$_1$) of a 3-hour transit of putative 1~\Re planet transiting each star in the LAMOST sample. S/N$_{1}$ was computed according to

\begin{equation}
\label{eqn:mes-formula}
\mathrm{S/N_1} = 
\left(\frac{\Rp}{\Rstar}\right)^{2}
\left(\frac{1}{\mathrm{CDPP3}}\right),
\end{equation}
where \Rp = 1~\Re and \Rstar was computed using LAMOST \teff, \logg, and \fe measurements and the publicly available {\tt isoclassify} package \citep{Huber17}.%
\footnote{https://github.com/danxhuber/isoclassify}
As Figure~\ref{fig:smet-detect} shows, we do not observe a significant dependence of planet detectability with stellar metallicity.

\clearpage
\section{Planet Occurrence: Period and Radius}
Here we provide two tables to supplement Section~\ref{sec:occur-per-radius} which treats planet occurrence as a function of period and radius. Table~\ref{tab:occurrence} lists the occurrence measurements displayed in Figure~\ref{fig:checkerboard} along with upper limits. Table~\ref{tab:population} is a sampling of the occurrence distribution shown in Figure~\ref{fig:contour}. We sample the occurrence the where the following conditions are met:
\begin{enumerate}
\item $P$ = 1--300~days (i.e., the range of periods shown in Figure~\ref{fig:contour}).
\item $\Rp$ = 0.5--32~\Re (i.e., the range of planet sizes shown in Figure~\ref{fig:contour}).
\item $\log\left(\Rp/\Re\right) > 0.15\, \log \left(P/1\, \mathrm{day}\right)$, which ensures that pipeline completeness, $\pdet > 0.25$.
\end{enumerate}
The integrated occurrence within this domain is \samp{pop-intrate} planets per 100 stars. We simulated the periods and radii of a population of \samp{pop-nplanets} planets in a sample of \samp{pop-nstars} Sun-like stars by drawing \samp{pop-nplanets} ($P$,$\Rp$) pairs according to their measured occurrence rates using the Python package pinky. These samples are listed in Table~\ref{tab:population}.

\begin{deluxetable*}{lrrrrrrR}
\tablecaption{Planet Occurrence\label{tab:occurrence}}
\tablecolumns{7}
\tablehead{
\colhead{$R_{P,1}$} & 
\colhead{$R_{P,2}$} & 
\colhead{$P_1$} & 
\colhead{$P_2$} & 
\colhead{\npl} & 
\colhead{\pdet} & 
\colhead{\ntrial} & 
\colhead{\fcell}
}
\tablewidth{0pt}
\startdata
1.00 & 1.41 & 1.00 & 1.78 & 7 & 0.95 & 6551.6 & $0.12^{+0.05}_{-0.04}$ \\
1.00 & 1.41 & 1.78 & 3.16 & 15 & 0.93 & 4369.1 & $0.36^{+0.10}_{-0.08}$ \\
1.00 & 1.41 & 3.16 & 5.62 & 40 & 0.89 & 2874.8 & $1.41^{+0.23}_{-0.21}$ \\
1.00 & 1.41 & 5.62 & 10.00 & 48 & 0.84 & 1847.5 & $2.63^{+0.39}_{-0.36}$ \\
1.00 & 1.41 & 10.00 & 17.78 & 51 & 0.76 & 1144.3 & $4.50^{+0.64}_{-0.59}$ \\
1.00 & 1.41 & 17.78 & 31.62 & 23 & 0.66 & 673.5 & $3.50^{+0.76}_{-0.65}$ \\
1.00 & 1.41 & 31.62 & 56.23 & 10 & 0.53 & 370.5 & $2.87^{+0.95}_{-0.78}$ \\
1.00 & 1.41 & 56.23 & 100.00 & 4 & 0.39 & 187.4 & $2.49^{+1.26}_{-0.97}$ \\
1.00 & 1.41 & 100.00 & 177.83 & 0 & 0.27 & 86.4 & $< 2.61$ \\
1.00 & 1.41 & 177.83 & 316.23 & 0 & 0.16 & 36.3 & $\nodata$ \\
\enddata
\tablecomments{Planet occurrence computed over various intervals in the $P$--\Rp plane spanning 0.25 dex and 0.15~dex, respectively. Each bin boundary is given by [$R_{P,1}$,$R_{P,2}$] and [$P_1$,$P_2$] respectively. We list the number of planets per bin \npl, the pipeline detectability \pdet, and the number of effective trials \ntrial. The number of planets per 100 stars per bin is given by \fcell, which is also shown graphically in Figure~\ref{fig:checkerboard}. We report 90\% upper limits on \fcell when there are no planets in a bin, and we do not report \fcell when $\pdet < 0.25$.  Table~\ref{tab:occurrence} is available in its entirety in machine-readable format. A portion is shown here for guidance regarding its form and content.}
\end{deluxetable*}

\begin{deluxetable*}{lRR}
\tablecaption{Simulated Planet Properties\label{tab:population}}
\tablecolumns{3}
\tablehead{
\colhead{Planet~~~~~~~~~~~~~~~~~~~~~~~~~} & 
\colhead{$P$} & 
\colhead{\Rp} \\
\colhead{~~~~~~~~~~~~~~~~~~~~~~~~~} &
\colhead{days} & 
\colhead{\Re}
}
\startdata
0 & 45.194 & 1.185  \\
1 & 94.327 & 0.964  \\
2 & 56.235 & 1.100  \\
3 & 44.367 & 2.254  \\
4 & 26.778 & 0.827  \\
5 & 67.644 & 6.516  \\
6 & 28.391 & 1.366  \\
7 & 236.817 & 7.830  \\
8 & 336.385 & 11.736  \\
9 & 22.677 & 2.457  \\
\enddata
\tablecomments{Simulated periods and radii of \samp{pop-nplanets} planets in a population of \samp{pop-nstars} stars based on the measured occurrence rates from \Kepler (Section~\ref{sec:occur-per-radius}). This table may be used to compute yield simulations for future surveys or integrated occurrence values over arbitrary bins of $P$ and $\Rp$. Table~\ref{tab:population} is available in its entirety in machine-readable format. A portion is shown here for guidance regarding its form and content.}
\end{deluxetable*}

\end{document}